\documentclass[12pt]{article}

\newcommand{\e}{\varepsilon}
\newcommand{\n}{\noindent}
\newcommand{\x}{\mathbf x}
\newcommand{\m}{$\mu\;$}
\usepackage{epsfig,amssymb,latexsym,color,amsmath,pifont}

\begin{document}

\title{Self-organization in Trees and Motifs of Two-Dimensional Chaotic Maps with Time Delay}
\author{Zoran Levnaji\'c and Bosiljka Tadi\'c}
\maketitle

\vspace{-0.35in}
\begin{center}
\it{Department of Theoretical Physics, Jo\v zef Stefan Institute \\
Jamova 39, SI-1000 Ljubljana, Slovenia}\\
\tt{zoran.levnajic@ijs.si, bosiljka.tadic@ijs.si} 
\end{center}

\begin{abstract}
\n We study  two-dimensional chaotic standard  maps  coupled along the edges of scale-free trees and tree-like subgraph (4-star)  with a non-symplectic coupling and time delay between the nodes. Apart from the chaotic and regular 2-periodic motion, the coupled map system exhibits variety of dynamical effects in a wide range of coupling strengths. This includes dynamical localization, emergent periodicity, and appearance of strange non-chaotic attractors. Near the strange attractors we find long-range correlations in the  intervals of return-times to specified parts of the phase space. We substantiate the analysis with  the finite-time Lyapunov stability. We also give some quantitative evidence of how the small-scale dynamics at 4-star motifs participates in the genesis of the collective behavior at the whole network.
\end{abstract}

\section{Introduction}
{\it Network Dynamics, Diffusion \& Chaos.}
Complex dynamical systems can be efficiently modeled by network topologies where the edges represent the interactions between the dynamical variables attached to the network nodes \cite{boccaletti,dorogovtsev,bosarev}. Various dynamical processes on the networks have been studied in which the temporal fluctuations in the dynamical variables constrained by the network structure are subject to the inter-node interactions and/or driving by the external noise. 
The interplay between the network topology and emergent dynamical behavior can be readily demonstrated in diffusive processes on scale-free networks \cite{bosarev}.  Of particular importance for the dynamics are  the networks  with well developed modular structures, topological subgraphs, which determine their own time scales \cite{arenas}. Such structures are often present in biological systems, in particular, in gene interaction networks, where structural modules, or dynamical motifs \cite{milo}, represent gene functional units mutually connected through the network. Although the cooperativity in gene functions can be easily demonstrated, e.g.,  with the appropriate statistical analysis \cite{jelena} of the empirical gene-expression  data, both the dynamics of gene motifs and the structure of gene networks remains a challenging problem. Recently, an application of discrete-time dynamics has been suggested for  modeling gene interactions \cite{schuster,lima,coutinho}. This approach brings a new prospective, in particular with respect to the advanced stability analysis and an intriguing  possibility of a chaotic behavior. 

The coupled chaotic maps on regular networks \cite{baroni,ahlers,zahera}, structured small graphs \cite{pisarchik,altmann,rama} and scale-free networks \cite{manrubia,lind,willeboordse,ljupco,ja3} are currently the subject of an intensive study. Many of the known one-dimensional chaotic maps were studied with the emphasis on the effects of coupling strengths  and of the time delay \cite{atay,masoller,kurths} between the units. The central points in these studies are the phenomena related with the synchronization of spatially extended coupled maps \cite{ahlers,rama,manrubia,ljupco,atay,masoller,kurths}, nature of the transition to the synchronized state \cite{baroni} and coexistence of different types of attractors \cite{pisarchik}. 

Generally, coupled chaotic maps are interesting dynamical systems in which the  dynamic stability can be examined with respect to variations in: 
\begin{itemize}
\item{} Initial conditions (the trajectory divergence); 
\item{} Parameters (coupling at a node, coupling between the units, geometry, time delay);
\item{} Drivings (external frequencies, endogenous driving).
\end{itemize} 
In this respect, two-dimensional coupled chaotic maps, which are much less studied so far, leave more room for the emergent dynamical effects \cite{ahlers,ja3,altmann}. In particular, in a system of standard maps with symplectic coupling on a full graphs trapping of the trajectories and anomalous diffusion was found \cite{altmann}. In the coupling with time delay on scale-free tree the standard maps lead to characteristic dynamical patterns  with robust statistical properties \cite{ja3}.  Type of the coupling between two degrees of freedom for a general one-dimensional map  coupled along a linear chain was shown 
\cite{ahlers} to affect the transition to the synchronized state.

In this work we study the discrete versions of the two-dimensional periodically kicked oscillators, usually referred to as the Chirikov standard maps \cite{chirikov}, coupled along their phase variables on large scale-free trees and their typical subgraph, the 4-star motif. As described in detail below, we impose a fixed {\it time delay} between coupled neighboring units (maps) thus rendering the coupled maps system {\it non-symplectic}. Our main purpose is to characterize the dynamic behavior that emerges as a consequence of the inter-node coupling both at small scale structure - the 4-star motif, and at the large scale structure - the tree with $N=1000$ nodes. Between the chaotic phase, which persists at weak coupling, and the fully synchronized phase with 2-periodic orbits, which dominates when nodes are strongly coupled, we find a whole range of couplings with a self-organized behaviors in which nodes influence each other. In this region the dynamical orbits are generally non-chaotic, with different patterns of collective dynamic behaviors, that we devote most of our attention to in this work. 

The paper is organized as follows. After defining our coupled map system in Section \ref{The Graphs of Coupled Maps}, we report the qualitative and statistical features of the coupled maps in Section \ref{Orbits Structure and Properties}. In Section \ref{Dynamics Stability} we investigate the dynamic stability of system's emergent orbits and describe behavior near the fractal attractors. Finally in Section \ref{Conclusions} we give a short summary and discussion of the results.

\section{The Graphs of Coupled Maps}\label{The Graphs of Coupled Maps}
We consider the standard map \cite{chirikov} in its original unbounded version, with $x$ being the phase (angle) and $y$ being the action (momentum) variable defined on each node $[i]$ of a network with $N$ nodes: 
\begin{equation} \left( \begin{array}{c}
x'[i] \\
y'[i]
\end{array} \right) = \left(
\begin{array}{l} 
  x[i] + y[i] + \e \sin (2 \pi x[i])  \;\;\;  \mbox{mod} \; 1  \\
  y[i] + \e \sin (2 \pi x[i])
\end{array} \right). \label{sm} \end{equation}
This map has been extensively studied as an \textit{isolated} system and its properties are well known, in particular the ones regarding the phase space diffusion for $\e$-values above the chaotic transition threshold ($\e_c \sim 0.155$). This is a symplectic map (a discrete version of a Hamiltonian system) which has a very intricate phase space structure with invariant KAM-torai persisting below the $\e_c$ value \cite{ll}.

We define our Coupled Map System (CMS) in accordance with the usually adopted scheme \cite{lind} as: $(1-\mu) \times (\mbox{standard}$ $\mbox{map-update}) \;$ + $\; \mu \times (\mbox{coupling})$. Also in analogy with the previous works on 1D maps \cite{zahera,lind,amritkar}, we consider the diffusive phase-coupling in the $x$-variable of Eq.\,(\ref{sm}). A one-step time-delay between the neighboring nodes is imposed, so that our CMS reads: 
\begin{equation} 
\left(\begin{array}{l}
x[i]_{n+1} \\
y[i]_{n+1}
\end{array}\right)
=
(1- \mu) 
\left(\begin{array}{l}
x[i]_n' \\
y[i]_n'
\end{array}\right)
+
\frac{\mu}{k_i}
\left(\begin{array}{c}
\sum_{j \in {\mathcal K_i}} (x[j]_n - x[i]_n') \\ 
0
\end{array}\right).
\label{main-equation} \end{equation} 
Here, ($'$) denotes the next iterate of Eq.\,(\ref{sm}), $n$ is the global discrete time,  $\mu$ is the coupling strength, $[i]$ indexes the nodes on the graph, $k_i$ is node degree and ${\mathcal K_i}$ denotes the neighborhood of the node $[i]$. 
The update of each node is the sum of a contribution given by its standard map-update (the $'$ part) plus a coupling contribution given by the sum of differences between the node's phase-value and the phase-values of neighboring nodes in the previous iteration, normalized by the node's degree. 

In contrast to the well-studied phase-coupled 1D oscillators \cite{lind,amritkar,zahera}, our CMS defined by the Eqs.\,(\ref{sm}-\ref{main-equation}) is essentially a network of time-delayed phase-coupled 1D oscillators with an additional  coupling to the local momentum variable at each node. From a continuous-time viewpoint, seeing $x'(i,t) \equiv x[i]'(t)$ as a continuous variable over a discrete space-time defined by the network, this coupling functional form can be seen as:
\begin{equation} \begin{array}{l}
x'(i,t+1) = 
(1- \mu)x[i]' + \frac{\mu}{k_i} \left[ \sum_{j \in {\mathcal K_i}} (x[j]' - x[i]')  + \sum_{j \in {\mathcal K_i}} (x[j] - x[j]') \right] = \\
(1- \mu)x[i]' + \frac{\mu}{k_i} \left[ \nabla^2 x[i]'  - \sum_{j \in {\mathcal K_i}} \partial_t x[j]' \right]
\simeq  \left[(1- \mu) + \frac{\mu}{k_i} (\nabla^2 - k_i \partial_t) \right] x'(i,t)
\end{array} \label{continuous-version} \end{equation}
which represents a sort of diffusive phase-coupling on network. In comparison to the diffusive CMS that are  typically studied on chains (e.g. \cite{ahlers}), our model involves 2D elements on a network with a time-delay, causing additional dynamical effects. 
Note also that, in contrast to the study in Ref.\ \cite{altmann}, our coupling scheme defined by the Eq.\,(\ref{main-equation}) is non-symplectic and in addition it includes a fixed time-delay. This type of coupling is motivated by a phenomenological model of the node interaction coming from the oscillatory origin of the standard map. 
Another important feature of the CMS defined by Eq.\ (\ref{main-equation}) is its
systematic inhibition of the diffusion along the action coordinate, which is made possible by the factor  $(1-\mu)$ in front of $y[i]^\prime$ 
(see section \ref{Orbits Structure and Properties}.), thus allowing more coherent interactions among the coupled nodes.

\begin{figure}[!hbt]
\begin{center}
$\begin{array}{cc}
\includegraphics[height=2.2in,width=2.45in]{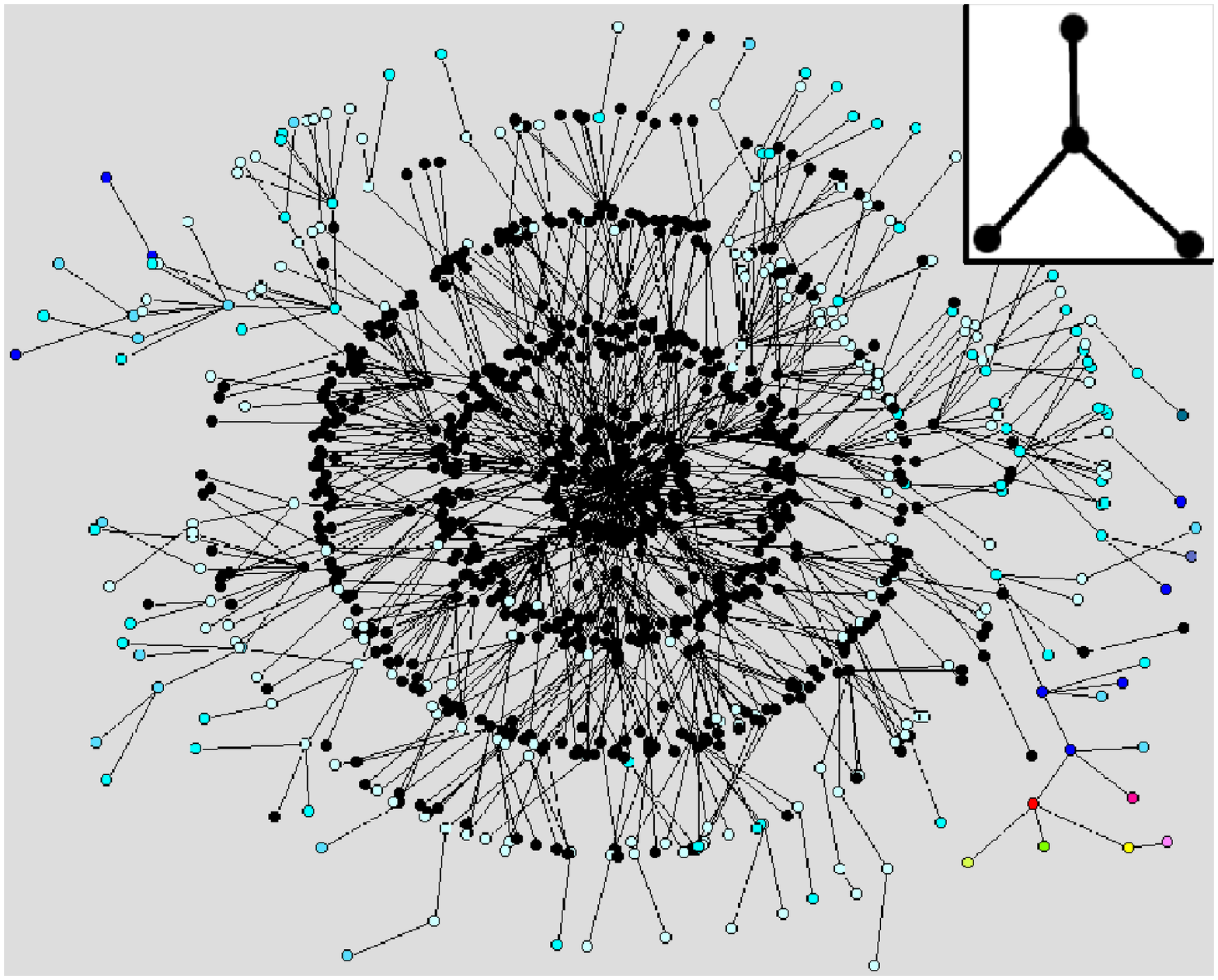} & 
\includegraphics[height=2.42in,width=2.6in]{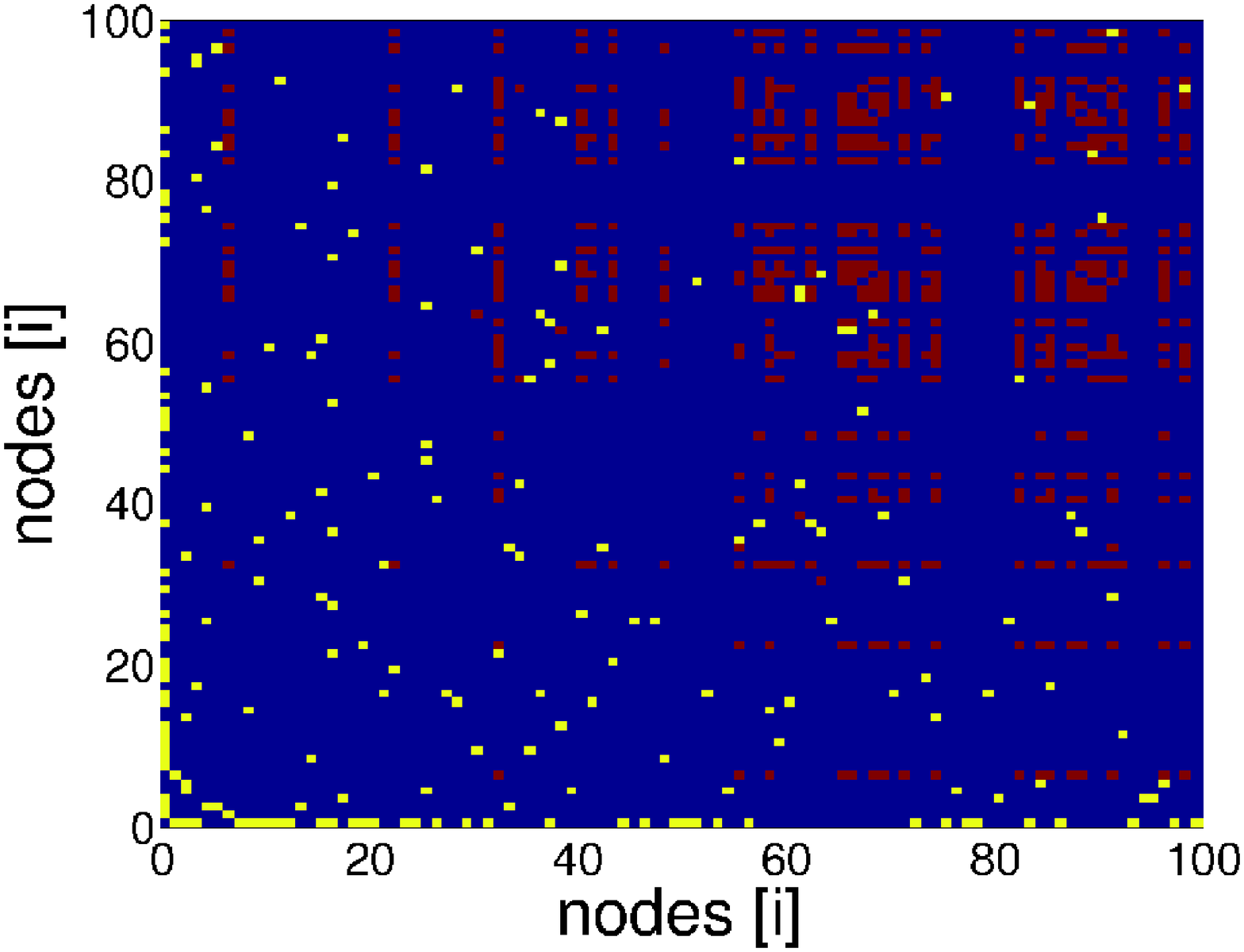} \\
\mbox{(a)} & \mbox{(b)} 
\end{array}$ 
\caption{(a) Scale-free tree at the threshold of regular dynamics with a pattern of nodes (bright color) which first achieve regular orbits; Inset: 4-star motif used in this work. (b) Pattern of connections (yellow) and pattern of synchronized nodes (red) on a 100-node scale-free tree for large inter-node coupling defined by Eq.\ (\ref{main-equation}).} \label{intro}
\end{center}
\end{figure}

{\it Geometries.} The CMS of Eq.\,(\ref{main-equation}) interact along the links of a network determined by its adjacency matrix $C_{ij}$. In this work we primarily focus on the coupling strength effects. Therefore, we consider tree-like structures in  order to avoid possible loop effects on the time-delayed interactions. In particular, we study CMS defined by Eq.\,(\ref{main-equation}) on a large scale-free tree with $N=1000$ nodes and on a typical  4-star motif, shown in Fig.\,\ref{intro}. Our previous study  on trees \cite{ja3} suggests that the chaotic maps with the coupling in Eq.\,(\ref{main-equation}) stabilize through the localization of orbits into well defined bands in the action coordinate. These bands determine different clusters of periodically synchronized nodes, which make a statistically stable pattern on the tree with the characteristic distance $d=2$ between two synchronized nodes \cite{ja3}.  Note that the 4-star is a typical motif which captures both the characteristic distance of synchronization and the tree-like nature of the topology. Here we systematically compare the emergent dynamic behavior of a 4-star and the whole tree in an extended range of couplings, where the CMS exhibits
new dynamical phenomena, as we describe below.

In Fig.\,\ref{intro}a  we show a typical situation of the tree at the threshold of localization as the coupling $\mu$ 
approaches a critical value $\mu_c\approx 0.021$ from below. In colors are shown nodes which first achieve periodic orbits, in contrast to black nodes, at which the maps still remain chaotic. The figure shows that the localization of orbits on the tree starts from the nodes with least number of links. In Fig.\,\ref{intro}b we show the adjacency matrix  and the situation with fully synchronized  nodes that occurs at large coupling strength $\mu = 0.08$. (For clarity a smaller scale-free network was shown.)  An interesting feature is that the pattern of synchronized nodes shows the largest density at the top-right corner, representing nodes at the boundary of the graph, whereas the pattern of direct links between nodes has largest density at the origin (near the hub node).  The results are collected from and averaged over many initial conditions.

In the following two sections we describe different types of dynamical behaviors obtained at different coupling strengths. Our particular emphasis is on the region $0.021 < \mu < 0.08$, i.e.,  between the above mentioned initial localization of orbits and the synchronization of nodes at strong couplings. We focus on the statistical properties of the collective motion and on the dynamic stability of emergent orbits. Throughout this work we fix the coupling $\epsilon =0.9$ between two components $(x[i],y[i])$ at each node  and the time delay $\tau =1$ between coupled nodes and we vary the inter-node coupling $\mu$. The results are collected  after an initial transient, typically 100000 steps for each trajectory, and averaged over many (typically 1000) initial conditions.  In all runs we pick the initial condition for all the nodes of the network randomly with a uniform probability from the phase space subset $(x,y) \in [0,1] \times [-1,1]$.

 \section{Orbits Structure and Dynamical Regimes }\label{Orbits Structure and Properties}

To study CMS on networks, the following types of  {\it emergent orbits} (e.o.)
can be considered: 
\begin{itemize}
\item{} Orbits of each individual node of the CMS, defined as
 \begin{equation} 
\left( x[i]_n,y[i]_n \right)_{n>n_0}, \;\;\; i=1,\cdots, N 
\label{eo}\end{equation}
where $n_0$ denotes the end of transients. For a selected node on the network we can in this way investigate in detail 
2D discrete dynamics using well known  study techniques of the dynamical systems.

\item{}Network-averaged emergent orbit (n.a.e.o.)  defined as
\begin{equation} 
\left( \hat{x}_n,\hat{y}_n \right)_{n>n_0} = \frac{1}{N}\sum_{i=1}^{N} (x[i]_n,y[i]_n)_{n>n_0}.
\label{naeo}\end{equation}
is another useful measure of the collective dynamics of the entire network, in particular in the context of the  synchronization and stability of the CMS.

\item{}Time-averaged emergent orbit (t.a.e.o.), defined for each node as
\begin{equation} 
\left( \bar{x}[i],\bar{y}[i] \right) = \lim_{n \rightarrow \infty} \frac{1}{n-n_0}\sum_{k=n_0}^{n} (x[i]_k,y[i]_k),
\label{taeo}\end{equation}
reduces the whole emergent orbit of a node to a single  point in the phase space, that qualitatively captures the emergent motion. 
\end{itemize}
In the following we always refer to a particular type of orbit used.

{\it Structure of Emergent Orbits of Coupled Maps.} 
As stated above, we consider CMS defined by Eq.\,(\ref{main-equation}) with fixed $\epsilon =0.9$ at each node and the time-delay between interacting nodes $\tau=1$, and vary the coupling strength between nodes $\mu$. Here we concentrate on the dynamical orbits of a selected node on the 4-star motif and a node on the tree graph. 
\begin{figure}[!hbt]
\begin{center}
$\begin{array}{ccc}
\includegraphics[height=1.7in,width=1.75in]{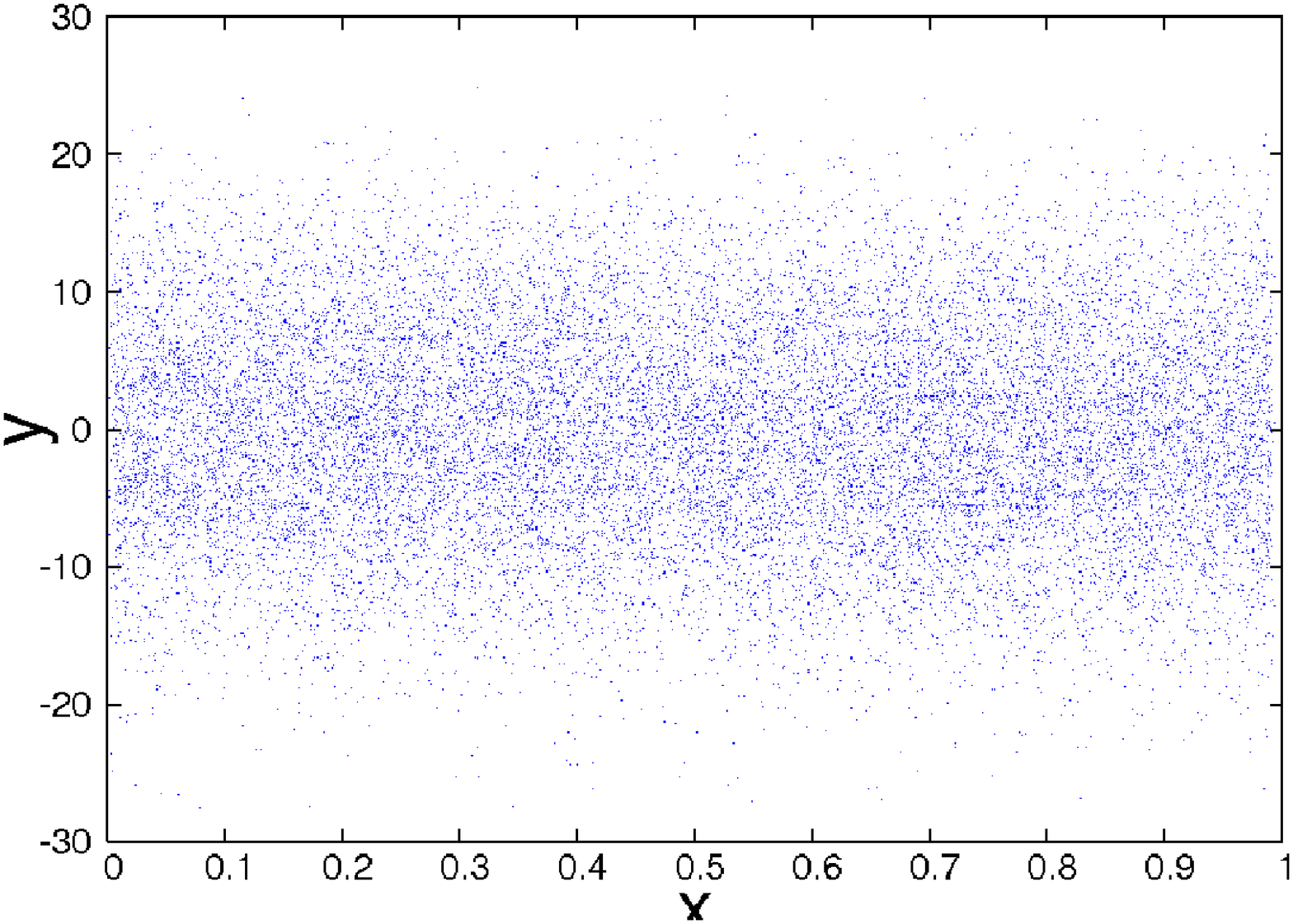} & 
\includegraphics[height=1.7in,width=1.75in]{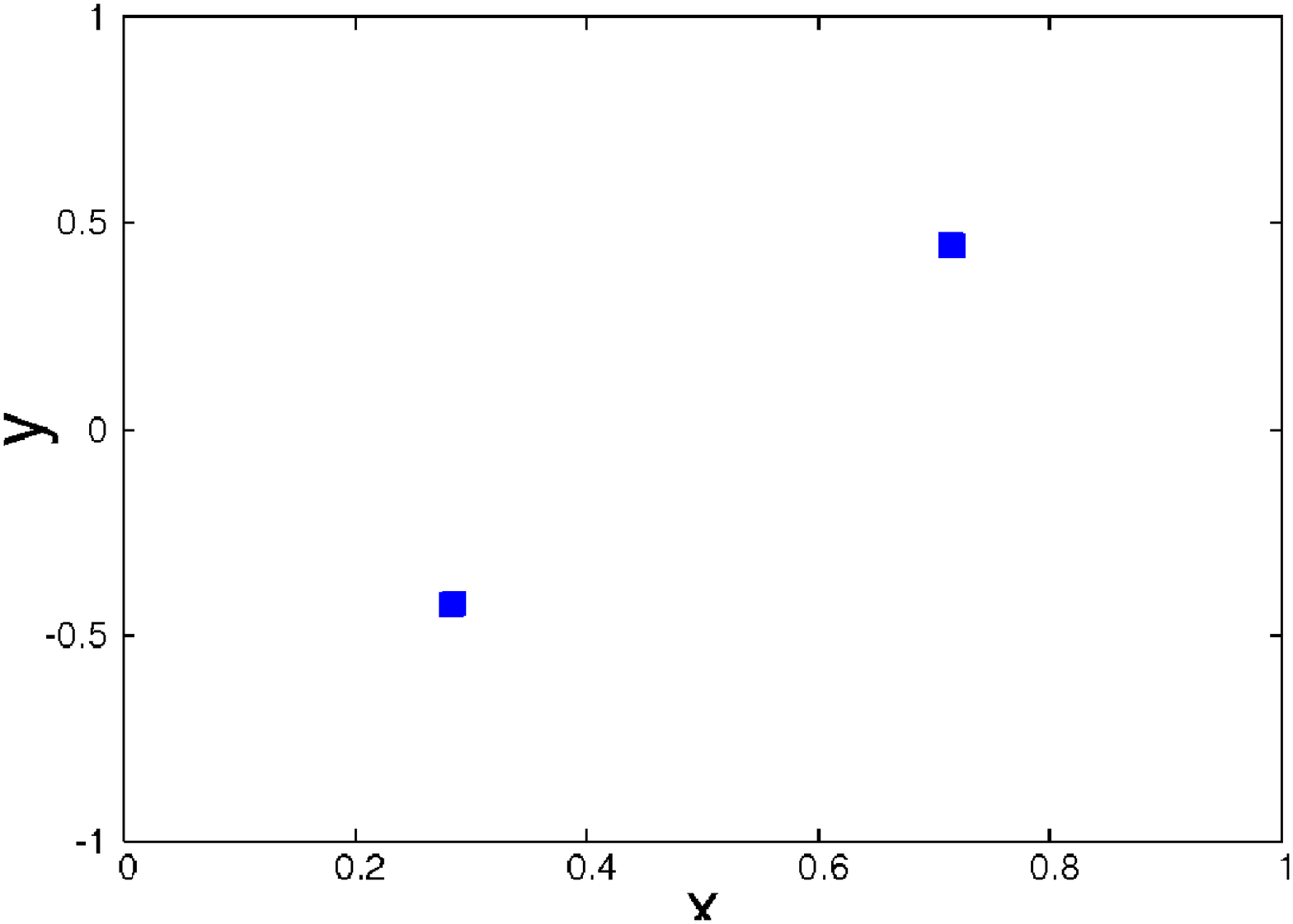} & 
\includegraphics[height=1.7in,width=1.75in]{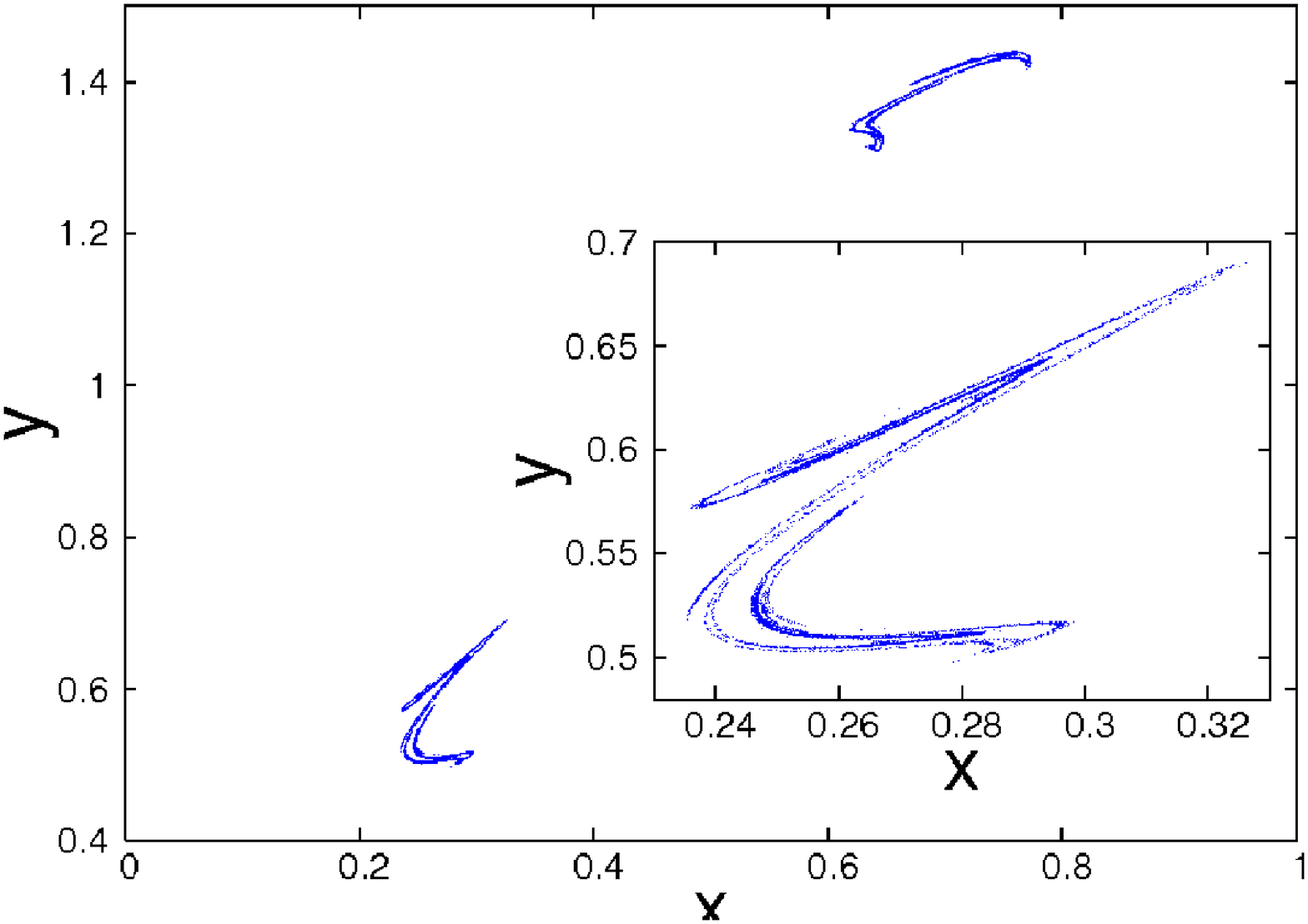} \\
\mbox{(a)} & \mbox{(b)} & \mbox{(c)} \\[0.2cm]
\includegraphics[height=1.7in,width=1.75in]{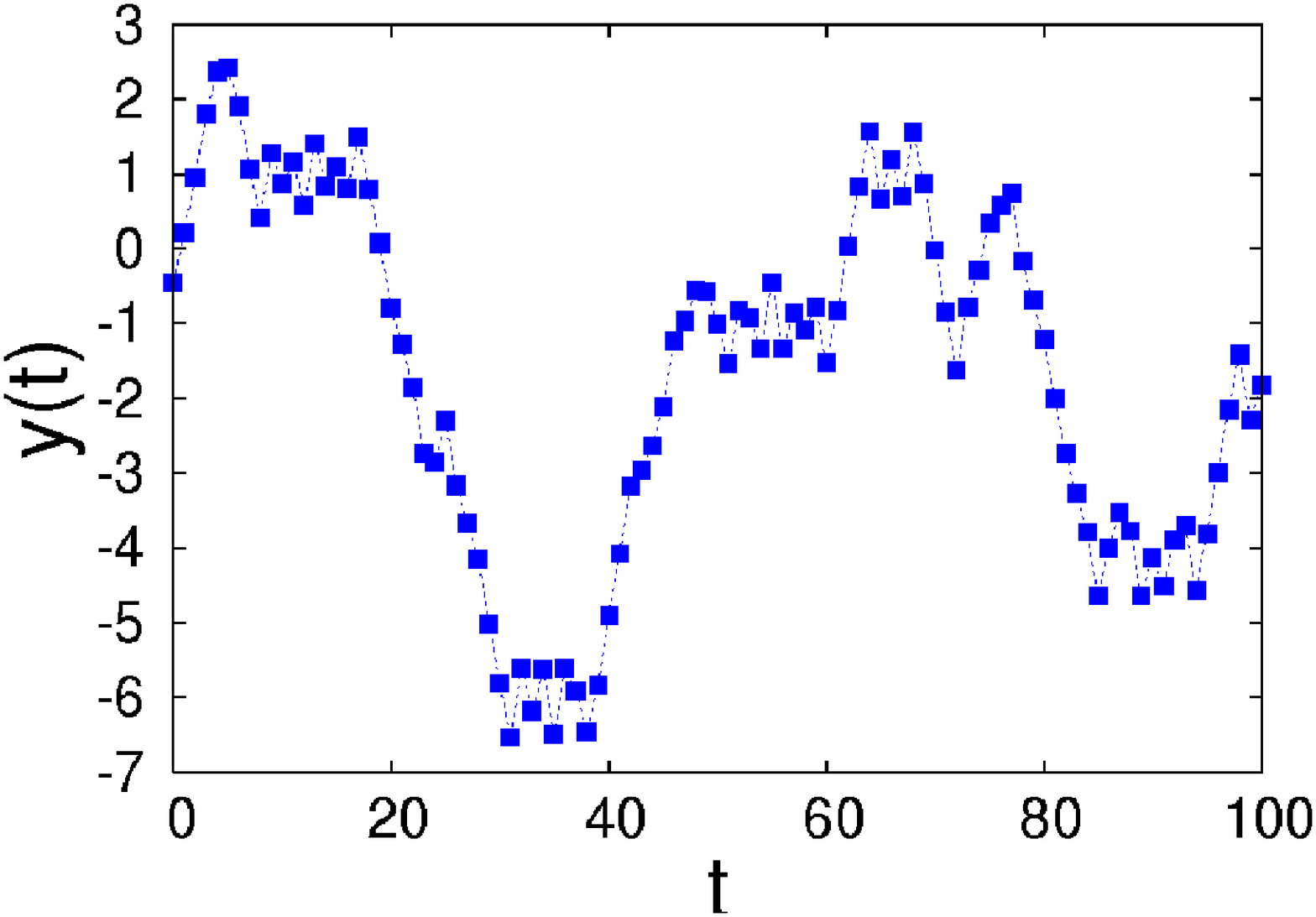} & 
\includegraphics[height=1.7in,width=1.75in]{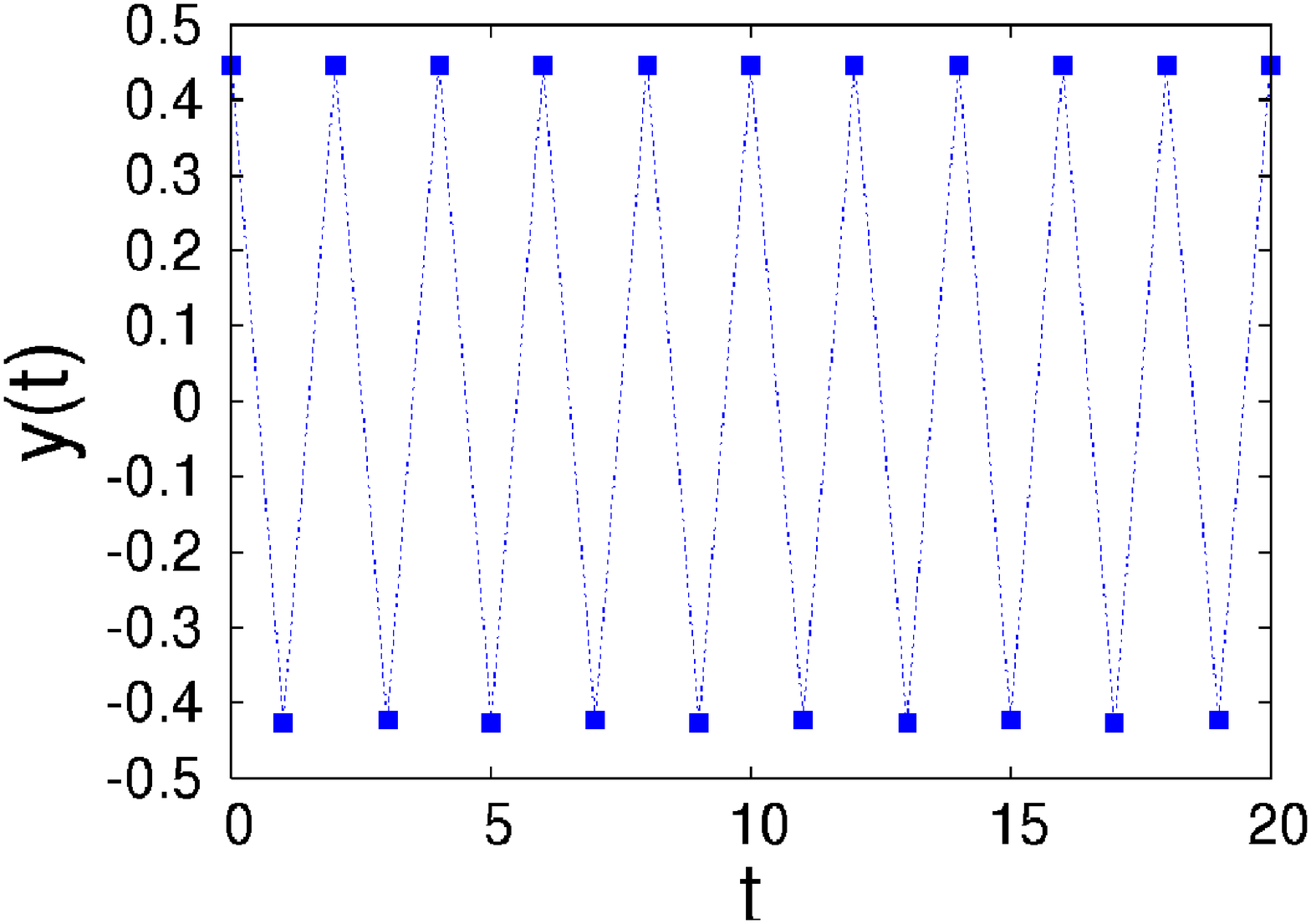} & 
\includegraphics[height=1.7in,width=1.75in]{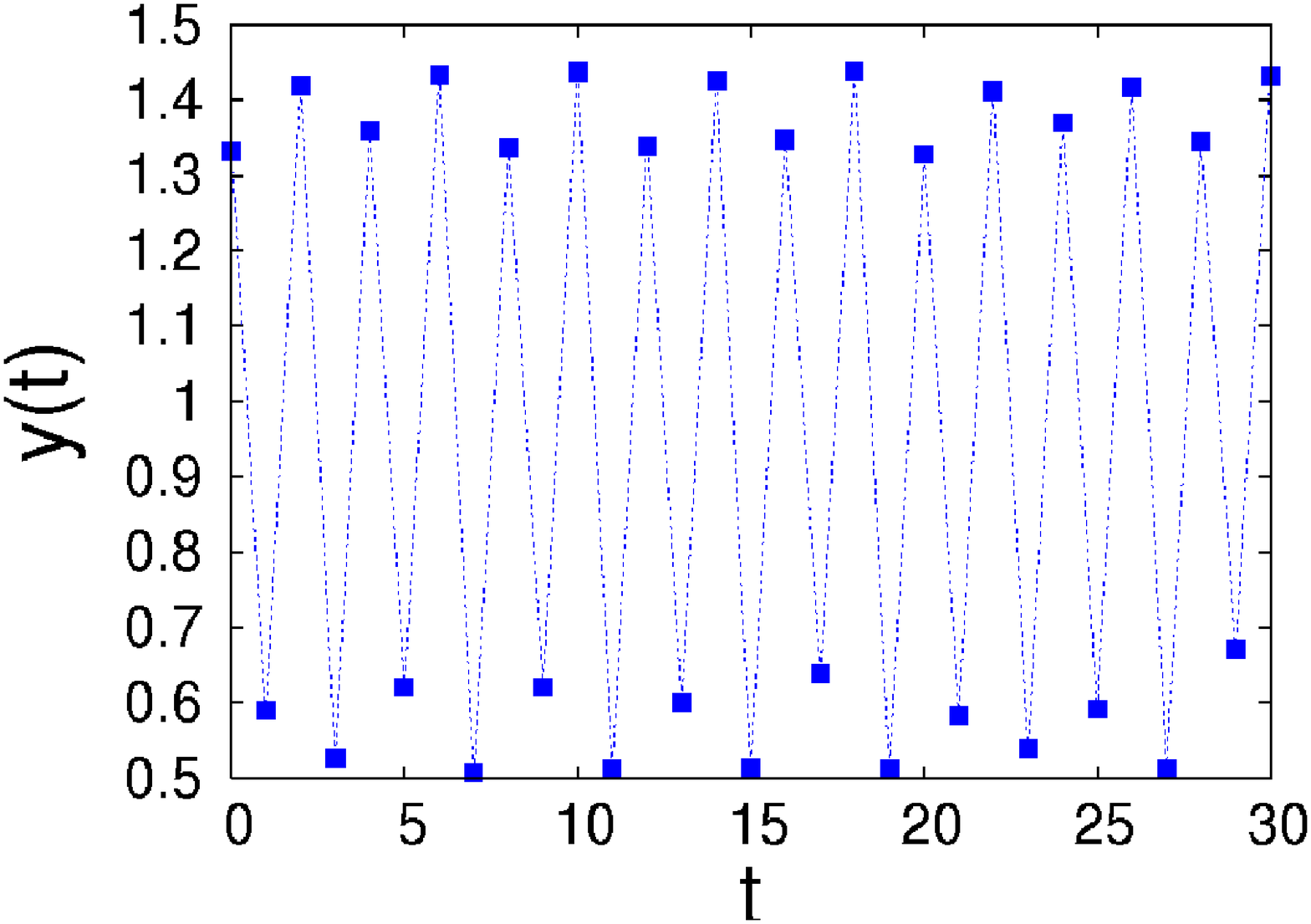} \\
\mbox{(d)} & \mbox{(e)} & \mbox{(f)} 
\end{array}$ 
\caption{Three typical emergent orbits (a--c) and the respective time-dependence of the $y$-coordinate (d--f) for a node on the 4-star motif: (a,d) chaotic orbit at $\mu=0.005$, and (b,e) periodic orbit at $\mu = 0.02$, both for the central node, and (c,f) strange attractor of a branch node at $\mu=0.049$. }
       \label{orbitsexamples}
\end{center}
\end{figure}

In Fig.\,\ref{orbitsexamples} we show  examples of three typical e.o. obtained for different \m-values for a node on the the 4-star motif. In particular, at weak coupling between nodes, $\mu=0.005$, we have a chaotic motion with orbits wandering in the phase space in an irregular fashion (Fig.\,\ref{orbitsexamples}a\&d), resembling the orbits of the uncoupled standard map Eq.\,(\ref{sm}). However, above a threshold coupling  $\mu \sim 0.02$ the CMS exhibits a regular motion. Orbits of different periodicity can be found in this region, such as the one in Fig.\,\ref{orbitsexamples}b. As a rule, the 2-periodic  orbits are the ones which survive at large couplings $\mu\ge 0.08$. In the region between $0.02 < \mu < 0.08$, depending on the initial conditions, we also find a non-periodic motion with the fractal attractors, as the example shown in  Fig\,\ref{orbitsexamples}c.  The time dependence of the momentum coordinate in the respective three dynamical regimes is also illustrated in Fig.\,\ref{orbitsexamples}d-f. Clearly,  our CMS of Eq.\ (\ref{main-equation}) exhibits a variety of emergent dynamical orbits,  in which both sensitivity to mutual couplings and to the initial conditions play a role. The appearance of the strange attractors with non-chaotic behaviors is a remarkable feature of the collective dynamics of our CMS, to which we  devote the Section \ref{Self-organized Strange Attractors}. Here we describe the differences between these three dynamical regimes in a quantitative manner.

\begin{figure}[!hbt]
\begin{center}
$\begin{array}{cc}
\includegraphics[height=2.03in,width=2.52in]{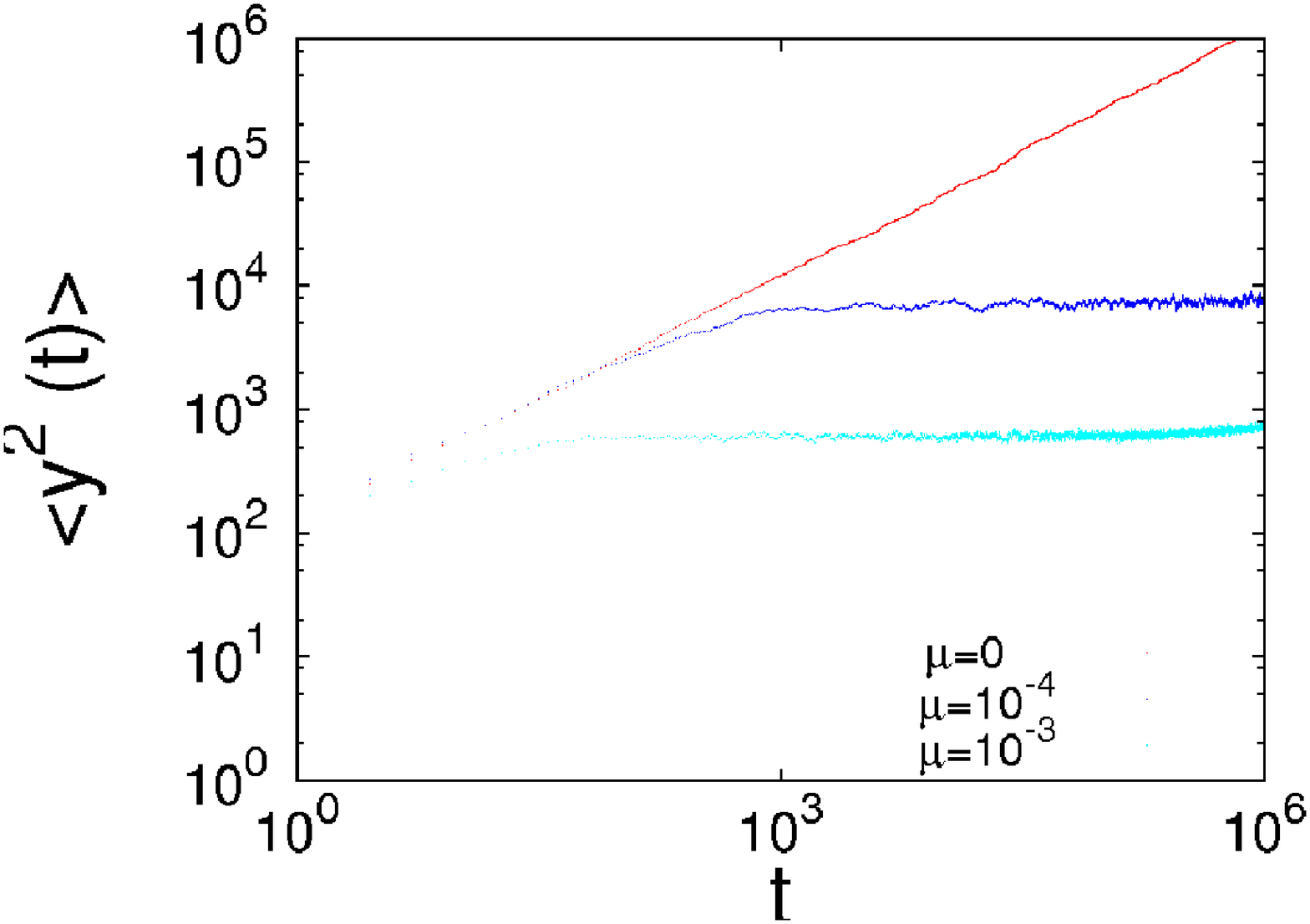} & 
\includegraphics[height=2.03in,width=2.52in]{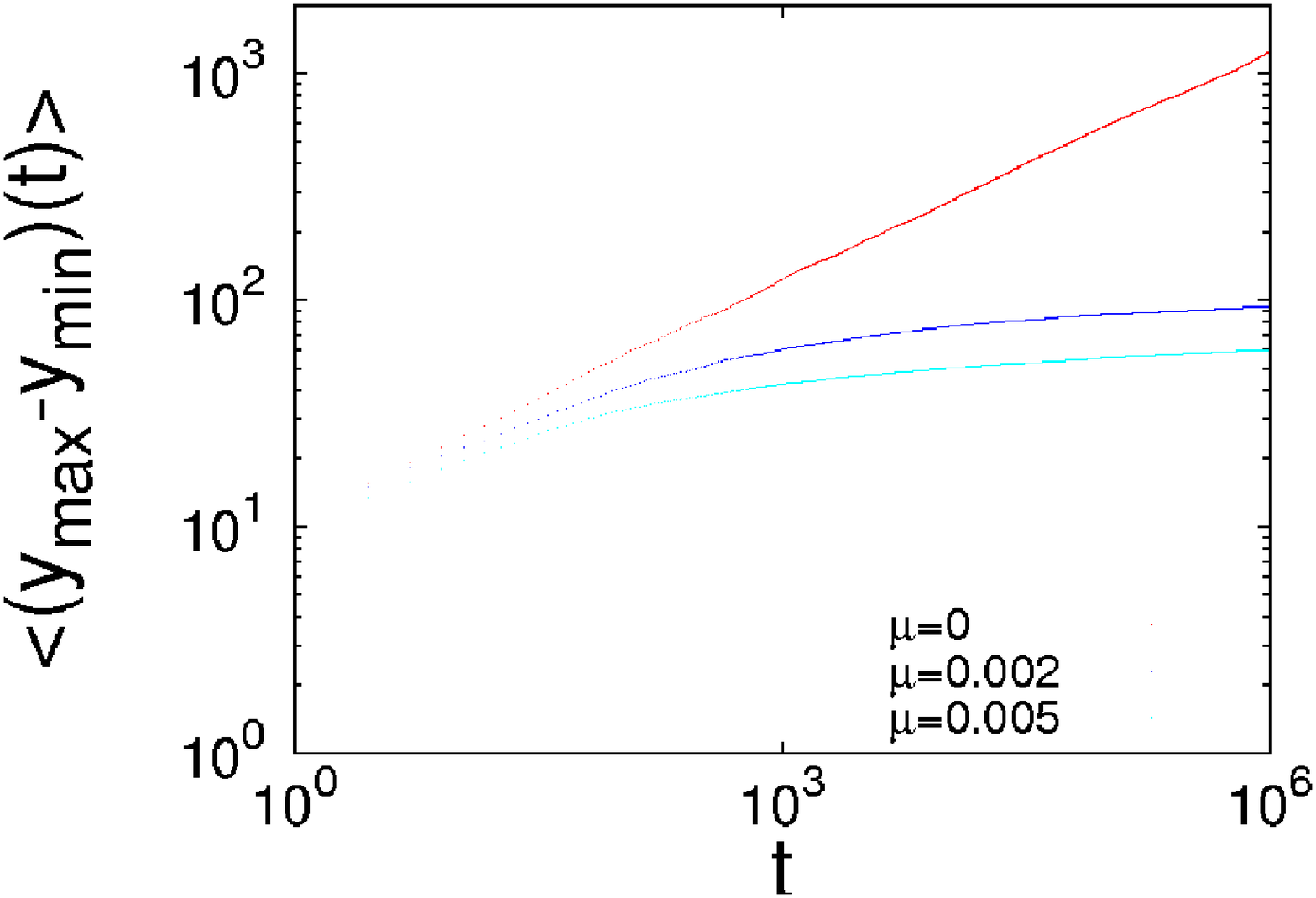} \\
\mbox{(a)} & \mbox{(b)} 
\end{array}$ 
\caption{Dynamical localization of the emergent orbits for various coupling strengths: $y$-coordinate mean square distance (a) and average stretching of the band width (b). Plots obtained by averaging over an ensamble of e.o. of 
4-star branch node. Transients were also included.}
       \label{diffusion}
\end{center}
\end{figure}

{\it Dynamical Localization.} 
In order to understand the nature of our CMS, we first analyze the chaotic e.o. occurring for arbitrarily small but non-zero couplings. Although these are chaotic orbits  (see also the stability analysis in the Section  \ref{Self-organized Strange Attractors}), in the presence of coupling they tend to localize  in the action $y-$coordinate, as shown in Fig.\,\ref{orbitsexamples}a\&d. Specifically,  as opposed to the chaotic diffusion in $y$-coordinate, which is  well-known for the uncoupled standard map Eq.\,(\ref{sm}) (see e.g. \cite{ll}),the interaction between nodes in Eq.\,(\ref{main-equation}) inhibits the diffusion. Quantitatively, the mean-square distance $<y^2(t)>$  as function of time $t$ for different coupling strengths $\mu$ is shown in Fig.\,\ref{diffusion}a. Clearly, the normal diffusion ($\gamma \simeq 1$) observed for uncoupled maps $\mu=0$ is replaced by the asymptotically localized orbits (finite m.s.r.) when $\mu > 0$ is turned on. A similar effect was demonstrated in the spreading of the band widths occupied by the e.o., Fig.\,\ref{diffusion}b. The band width in $y-$coordinate increases as a power of time for  uncoupled standard maps, whereas the stretching rate decreases  with a non-zero coupling. Apart from quantitative differences, such localization occurs both on large trees and on small motifs.

The observed dynamical localization is a clear argument that the collective effects due to node-interaction are occurring. As we shall see in the following paragraphs, further decrease in the  stretching rate of the band widths occupied by an e.o. triggers the appearance of a regular  motion.

\begin{figure}[!hbt]
\begin{center}
$\begin{array}{cc}
\includegraphics[height=2.1in,width=2.9in]{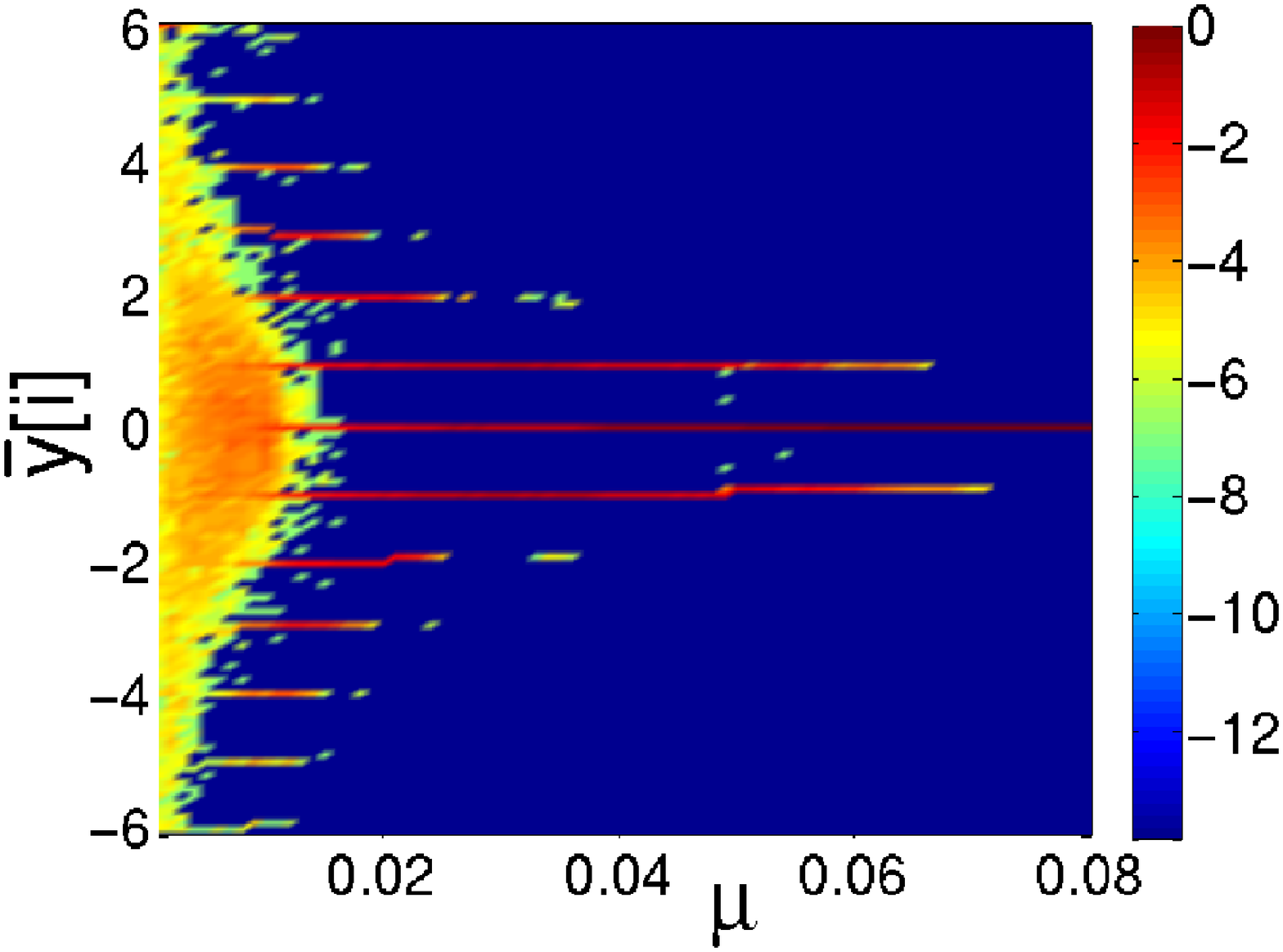} & 
\includegraphics[height=2in,width=2.1in]{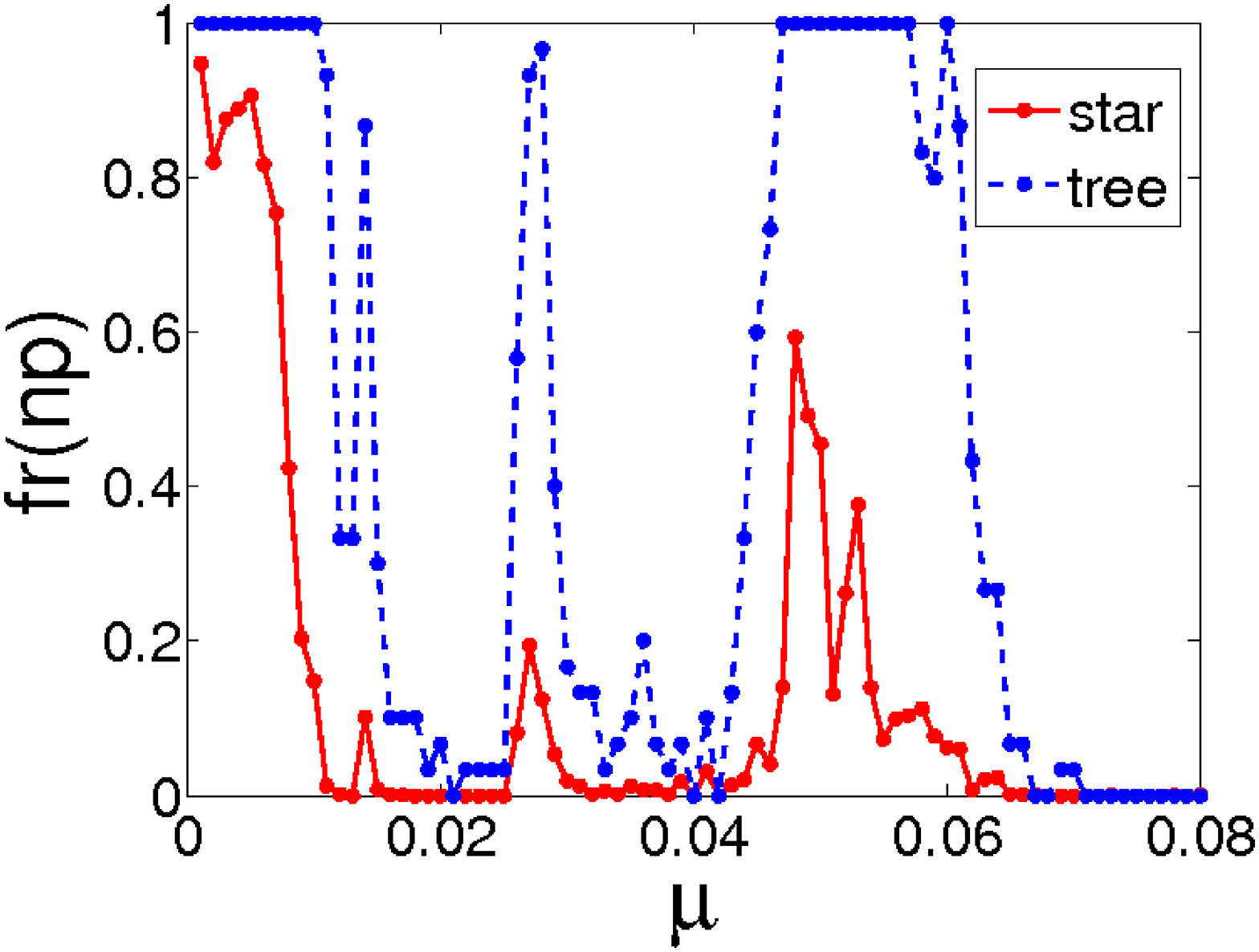} \\
\mbox{(a)} & \mbox{(b)} 
\end{array}$ 
\caption{(a) For a star-branch node: 2D distribution histogram of time-averaged emergent orbits $\bar{y}$ against coupling strength $\mu$. Color-code: logarithm of the fraction of e.o. corresponding to a given  $\bar{y}$. (b) Fraction of {\it non-periodic} e.o., $fr(np)$, as function of $\mu$ for a star's branch node and an outer node of the tree. Periods up to 10000 are considered.}
       \label{periodictransition}
\end{center}
\end{figure}

{\it Generalized Synchronization and Periodic Orbits.}
The appearance of periodic e.o. is one of the central collective feature exhibited by the CMS Eq.\,(\ref{main-equation}). The trajectories, similar to the one in Fig.\,\ref{orbitsexamples}b\&e,  oscillates in the phase space between two localized groups of points in a way to maintain a constant average $\bar{y}$. For further quantitative study we use the time-averaged e.o. $(\bar{x},\bar{y})[i]$ defined in Eq.\,(\ref{taeo}) and consider the star's branch-node. As t.a.e.o. is represented by a point in 2D node's phase space, here we plot only its $\bar{y}$-coordinate as function of the coupling strength \m, as shown in Fig.\,\ref{periodictransition}a. The color code represents (on log-scale) number of initial conditions that end up with a given value of $\bar{y}$.

The Gaussian distribution of $\bar{y}$-values for small $\mu$ reduces to an organized set of clusters as the coupling strength is increased above $\mu \simeq 0.01$. With the disappearance of the chaotic e.o., classes of periodic e.o. 
appear. In this region, depending on the initial conditions, an orbit ends up with a given $\bar{y}$-value, i.e., at one of the horizontal lines in Fig\,\ref{periodictransition}a. The e.o. perform an oscillatory motion around a fixed set of phase space points, which are  located symmetrically around  the respective average value $\bar{y}$. The emergent periodicity is studied in more details later. A similar structure of orbits is found for each node on the graph, with robust patterns of groups of nodes having  the same $\bar{y}$ value.  Spatial patterns of such clusters of nodes depend on the local structure of the graph \cite{ja3}. On the scale-free tree graph the number of possible groups gradually  decreases with increased coupling strength and eventually reduces to a single group centered around $\bar{y}=0$ for $\mu \simeq 0.07$.

{\it Appearance of Non-Periodic Orbits.} Furthermore, in the same region of couplings we find a scalar fraction of initial conditions leading to non-periodic orbits. In Fig.\,\ref{periodictransition}b we show the fraction of non-periodic e.o. 
against coupling strength both for a node on the 4-star and for a node on the scale-free tree. Nature of the non-periodic orbits in our CMS will be studied in more details in the next Section. Here we point out a systematic correlation between two curves in Fig.\,\ref{periodictransition}b, suggesting the importance of the  dynamics at small-scale, 4-star motif, in the genesis of the non-periodicity at the tree graph.

\subsection{ Statistical Features of Emergent Orbits}
As the above results and Fig.\,\ref{periodictransition} suggest, the coupled
chaotic maps  defined by Eqs.\ (\ref{sm}-\ref{main-equation}) on tree graphs
exhibit several types of collective dynamic behaviors, apart from synchronized regular motion, which dominates at large couplings. In particular, we find $(i)$ {\it weakly localized} chaos, $(ii)$ {\it emergent periodicity}, and $(iii)$ {\it non-periodic} orbits with non-chaotic features (see later). With the appropriate statistical analysis we further differentiate between these dynamical regimes.

\begin{figure}[!hbt]
\begin{center}
$\begin{array}{cc}
\includegraphics[height=2.1in,width=2.6in]{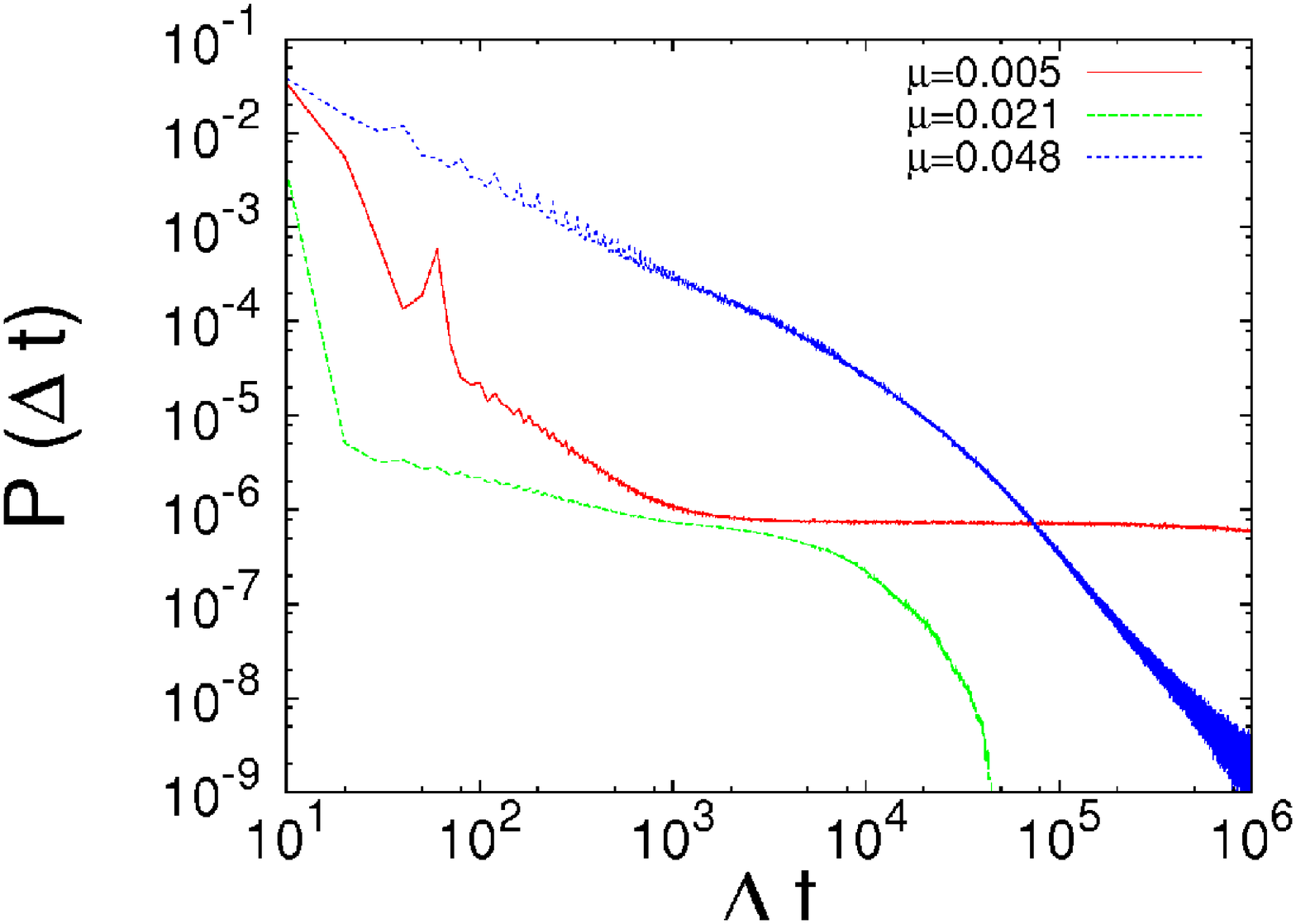} &
\includegraphics[height=2.0in,width=2.4in]{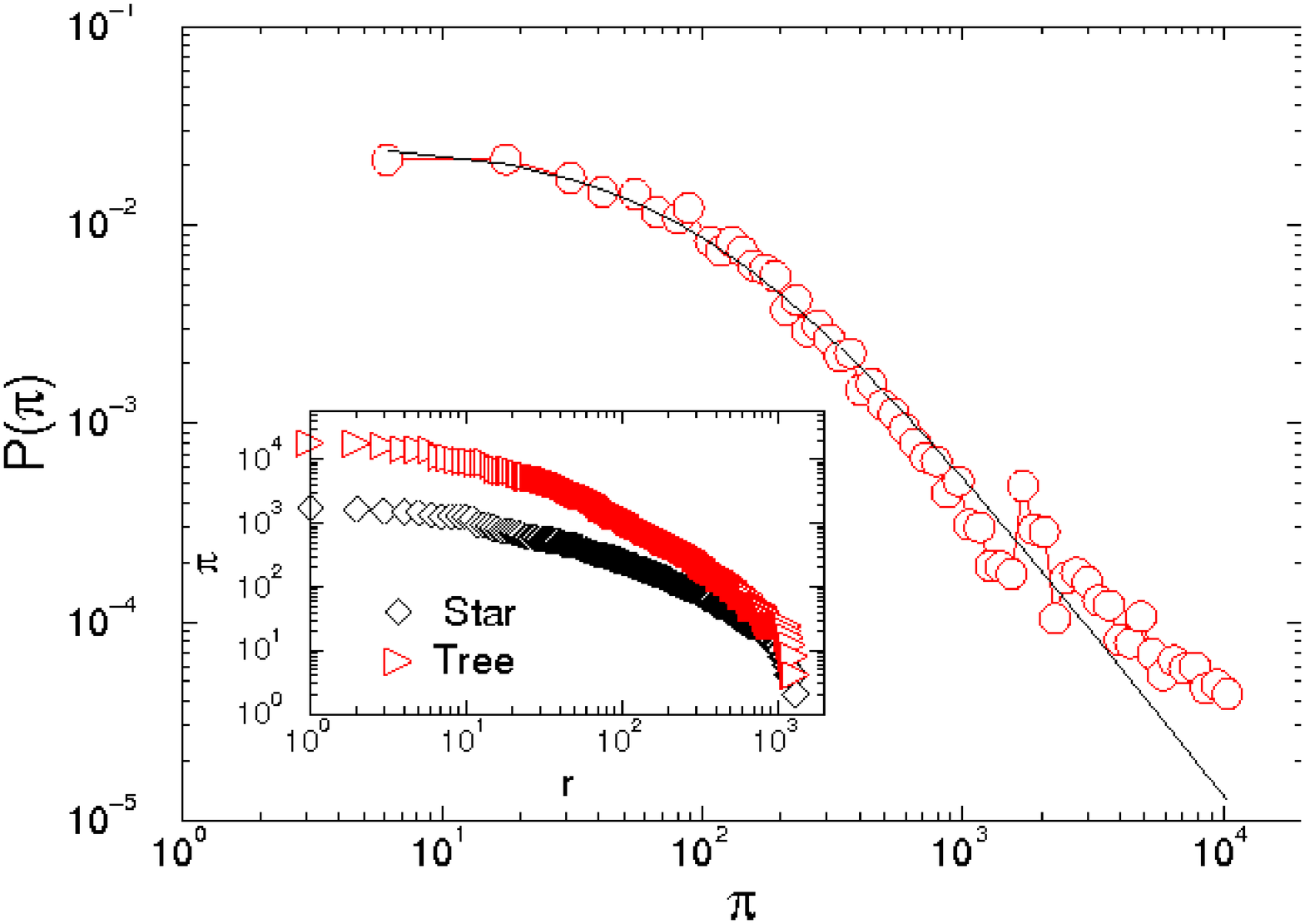} \\
\mbox{(a)} & \mbox{(b)} 
\end{array}$ 
\caption{  (a) Distribution of return-times to a selected part of the phase space in three dynamical regimes corresponding to three different coupling strengths \m for star's branch-node. Phase space grid of $1000 \times 100000$ covering $(x,y) \in [0,1] \times [-50,50]$ was used and distributions averaged over  an ensamble of initial conditions. (b) Distribution of periods of the network-averaged  orbits on the tree for $\mu=0.08$. Inset: Ranking statistics of periods for a branch node on the 4-star and an outer node on  the tree for $\mu =0.21$.}
       \label{statisticsperiodiceo}
\end{center}
\end{figure}

{\it Return-times statistics.} The first-passage (or return-time) statistics represents an interesting statistical measure of correlations in the dynamical systems. In this context, we monitor return of the trajectories to a marked small area in the phase space. The results, integrated over the phase space subset $(x,y) \in [0,1] \times [-50,50]$ and averaged over many initial conditions for three representative values of the coupling $\mu$ are shown in Fig.\,\ref{statisticsperiodiceo}a. Clearly, the distributions of return-times have different tails for the chaotic motion (the case with $\mu =0.005$), periodic (for $\mu=0.021$) and non-periodic  (the case with $\mu=0.048$) behavior. In particular, the cooperative behavior leading to the power-law tail of the distribution in the case for $\mu=0.048$, suggests that the observed non-periodic orbits in this regime are non-chaotic. It is interesting that the differences between these dynamical behaviors are also  expressed 
in the transient regime (approximately first $10^3$ iterations in Fig.\,\ref{periodictransition}a).    

{\it Distribution of Periods.} Another measure of the cooperative behavior in the CMS is the emergent periodicity
of the orbits at the level of an individual node coupled to the system and in the system-averaged trajectory.
In Fig.\,\ref{periodictransition}b we show the distribution of periods $\pi$ of the periodic orbits for the tree-average trajectory and, in the inset, ranking statistics of the periods observed at individual nodes coupled on a tree and on the 4-star. It is interesting that at the tree-like topologies the odd and even periodicities appear to have different statistics, which is especially pronounced at lower values of $\pi$ and at weaker couplings. Although the large periods can be found at nodes on small graphs, our findings suggest that the large-scale structures are more efficient in building-up the power-law dependencies in the distribution of periodicity. The network-averaged trajectories, defined by Eq.\ (\ref{naeo}), exhibit the emergent periodicities 
with the power-law distribution shown in Fig.\,\ref{periodictransition}b. The curve can be well fitted by the $q-$exponential form \cite{tsallis88} 

\begin{equation}
P(X) = B_q\left(1 -(1-q){{X}\over{X_0}}\right)^{{1}\over{1-q}} , \, 
\label{q-exp-form}
\end{equation}
with the characteristic period $X_0=63$ above which the power-law tail appears with the slope 1.66. The fit is compatible with the value of the parameter $q=1.6$. Often large deviations of the parameter $q$ from unity can be related with a non-ergodicity of the underlying phase space.

\section{Dynamic Stability and Evidence of SNA}\label{Dynamics Stability}
The analysis in previous Sections  revealed that our CMS maintains different types of cooperative dynamical behaviors, depending on the coupling strength. In order to further substantiate the nature of the emergent dynamics, in particular the occurrence of the non-periodic orbits, here we discuss the appropriate dynamic stability of the system. 

We analyze the trajectory divergence \cite{feudel} within small neighborhoods of the emergent orbit. As it was shown above, for the dissipative CMS of Eq.\,(\ref{main-equation}) the trajectories at long times tend to localize at a finite distance in the action coordinate. This localization implies that only {\it limited} divergence between different trajectories can be observed at $t\to \infty$. Therefore, the appropriate stability measure is the {\it initial divergence rate}, which lasts for a finite time interval and determines the finite-time Lyapunov exponent \cite{feudel} .

\subsection{Finite-Time Lyapunov Stability}

\begin{figure}[!hbt]
\begin{center}
$\begin{array}{cc}
\includegraphics[height=2.1in,width=2.21in]{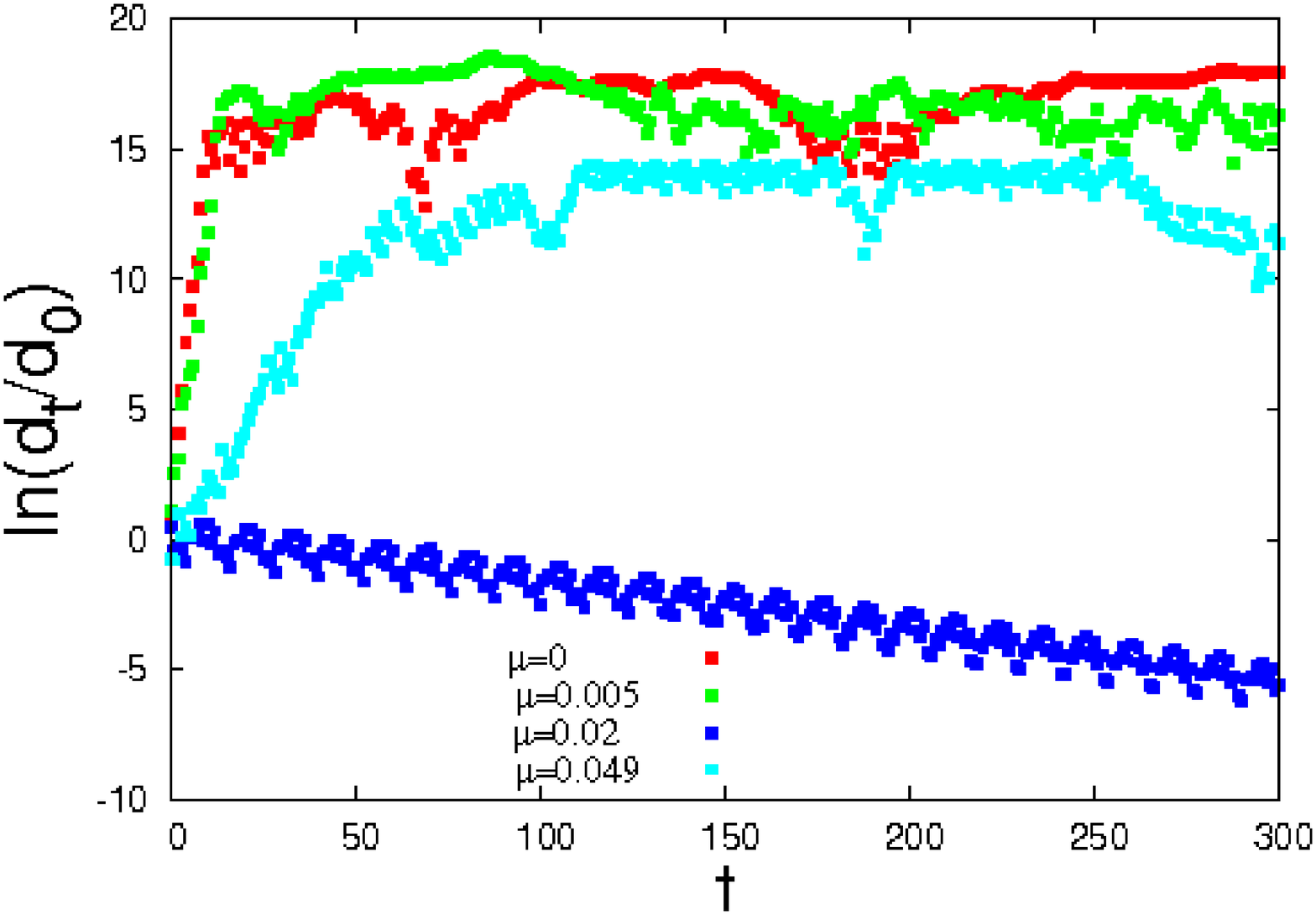} & 
\includegraphics[height=2.21in,width=2.9in]{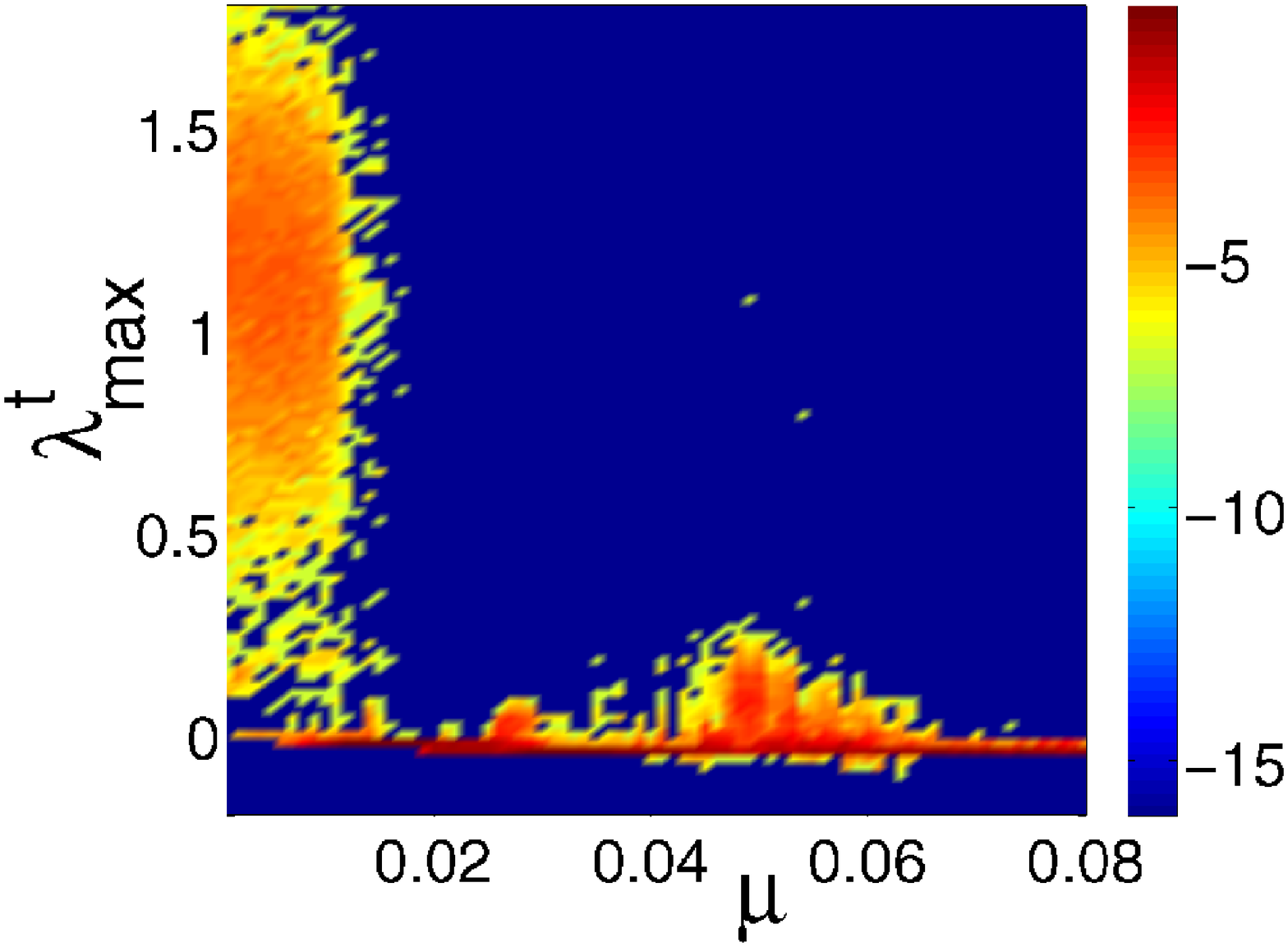} \\
\mbox{(a)} & \mbox{(b)} 
\end{array}$ 
\caption{(a) Behavior of the orbit divergence rates for four typical e.o. with different couplings \m. (b) 2D distribution histogram of $\lambda_{max}^t$ for star's branch-node versus \m. Color-code reports the logarithm of the fraction of orbits leading to the same $\lambda_{max}^t$ for a given \m-value.}
       \label{stabilityintro}
\end{center}
\end{figure}

The  Maximal Lyapunov Exponents (MLE) for a phase space point is generally defined as the maximum rate of the  divergence among all the trajectories starting in the vicinity of the considered point. 
In  Fig.\,\ref{stabilityintro}a we show several examples of different initial divergence rates for different trajectories. As expected, all  curves reach a saturation (slope zero) at long times. However,  the trajectory separation increases over certain number of initial iterations $n$ with a well defined slope, before the saturation occurs. These slopes determine the finite-time MLE,  denoted as $\Lambda_{max}^t$. 

More precisely, the finite-time Lyapunov exponent $\Lambda_{max}^t (\x_0)$ associated with the point $\x_0 \equiv (x_0,y_0)[i]$ on a trajectory is defined by
\begin{equation}
\Lambda_{max}^t (\x_0) = \max_{\x \in {\mathcal N}} \; \left\lbrace \mbox{initial slope}  \left[ \frac{1}{n}\ln 
\frac{d(U_n \x,U_n \x_0)}{d(\x,\x_0)} \right]\right\rbrace,
\label{mledefinition}
\end{equation}
where ${\mathcal N}$ stays for a small neighborhood around the point $\x_0$, $d(.,.)$ denotes distance  and $U_n$ 
stands for the discrete time-evolution dynamics given by CMS Eqs.\,(\ref{sm}-\ref{main-equation}). Note that $n$ should be appropriately selected for each trajectory separately. As before, we consider trajectories of a selected node coupled on the 4-star and on the tree.
 
The procedure  employed for computing the $\Lambda_{max}^t(\x_0)$ is the following:
\begin{enumerate}
\item take an initial condition $\x_0$ (for all the nodes) and compute their e.o.;
\item focus on the final state (point) of a selected node and consider a random point in its close neighborhood, 
with the  distance $d$ defined as ``Manhattan distance'' $d((x,y),(x_0,y_0))=|x-x_0|+|y-y_0|$; 
\item iterate the dynamics for both points with systematically recording the rate that their distance ratio Eq.\ (\ref{mledefinition}) changes in time and determine the optimal $n$;
\item approximate the slope over the optimal estimation time $n$. Repeat the same procedure for a few other points in the same neighborhood and determine the maximal slope obtained this way.
\end{enumerate}
Clearly,  $\Lambda_{max}^t (\x_0)$ corresponds to only one point on the e.o., implying that, using different points, a spectrum of  $\Lambda_{max}^t$-s can be determined to characterize the entire orbit. Furthermore, we define an average value $\lambda_{max}^t$ over the distribution of different $\Lambda_{max}^t (\x_0)$ for a given orbit  as:
\begin{equation} \lambda_{max}^t = ~<\Lambda_{max}^t (\x_0) >_{\x_0 \in \mbox{e.o.}} 
\label{lambde}\end{equation}

We compute the average $\lambda_{max}^t$ for a branch-node on the 4-star for many trajectories and different coupling strengths $\mu$. The results are shown in  Fig.\,\ref{stabilityintro}b in the form of 2D histogram for \m in  the interval [0.0,0.08]. Not surprisingly, fairly large positive values of $\lambda_{max}^t$ at $\mu < 0.01$ are in a full accordance with the chaotic motion of weakly coupled nodes. More interesting is the appearance of the orbits with $\lambda_{max}^t > 0$ in the range of couplings $0.01 < \mu < 0.08$, where most  of the trajectories are localized (cf. Fig.\,\ref{periodictransition}a\&b). Note, however, that the measured values of $\lambda_{max}^t$ in this region are far smaller compared to the chaotic e.o. at small \m. This indicates that the nature of these e.o. might be fundamentally different. It should be also noted that, in this region of couplings we can find both positive and negative  $\lambda_{max}^t$ for different initial conditions (or different parts of the trajectory), as discussed later. Qualitatively similar picture is found for the trajectories of an outer-node on the large tree.

\begin{figure}[!hbt]
\begin{center}
$\begin{array}{cc}
\includegraphics[height=2.05in,width=2.4in]{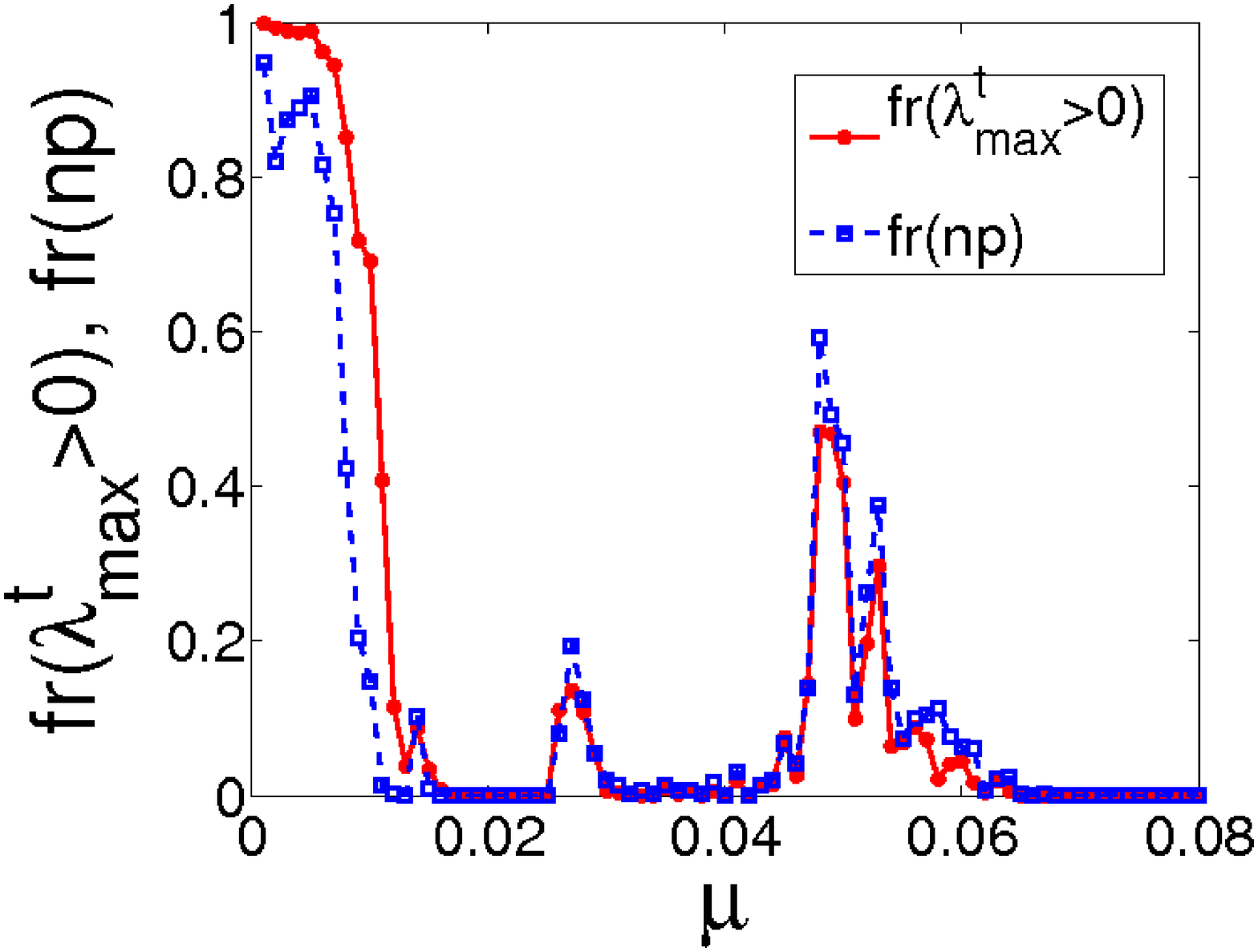} & 
\includegraphics[height=2.05in,width=2.4in]{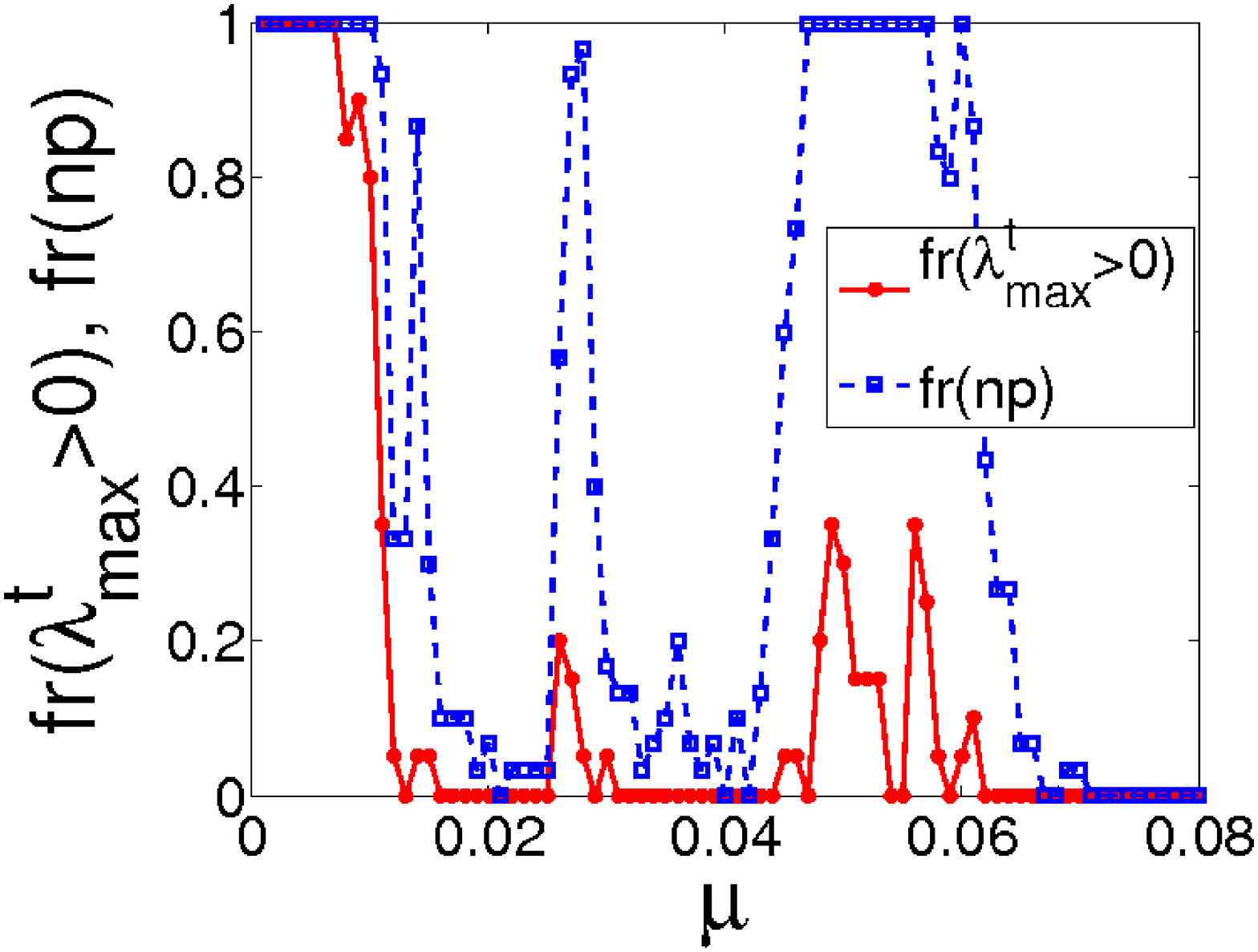} \\ 
\mbox{(a)} & \mbox{(b)} 
\end{array}$ 
\caption{ Comparison of the fraction of orbits with a positive $\lambda_{max}^t$ and fraction of non-periodic e.o. for  (a) a branch-node on the 4-star motif and  (b) an outer node on the large scale-free tree. }
       \label{comparisons}
\end{center}
\end{figure}
Next we compare the occurrence of the trajectories with $\lambda_{max}^t >0$ and the non-periodic orbits, mentioned before. Note that the periodicity is usually determined after long times when the trajectory is localized, whereas the value of $\lambda_{max}^t$ is the average computed along the entire trajectory. Nevertheless, as shown in Fig.\,\ref{comparisons}a\&b, we find a significant overlap in the fraction of the non-periodic orbits and the orbits with a positive finite-time Lyapunov exponent $\lambda_{max}^t$. In the case of a 4-star's node the overlap is nearly complete, whereas in the case of a tree's node the qualitative similarities are clear.

\subsection{Appearance of Strange Attractors}\label{Self-organized Strange Attractors}
\begin{figure}[!hbt]
\begin{center}
$\begin{array}{cc}
\includegraphics[height=1.7in,width=2.in]{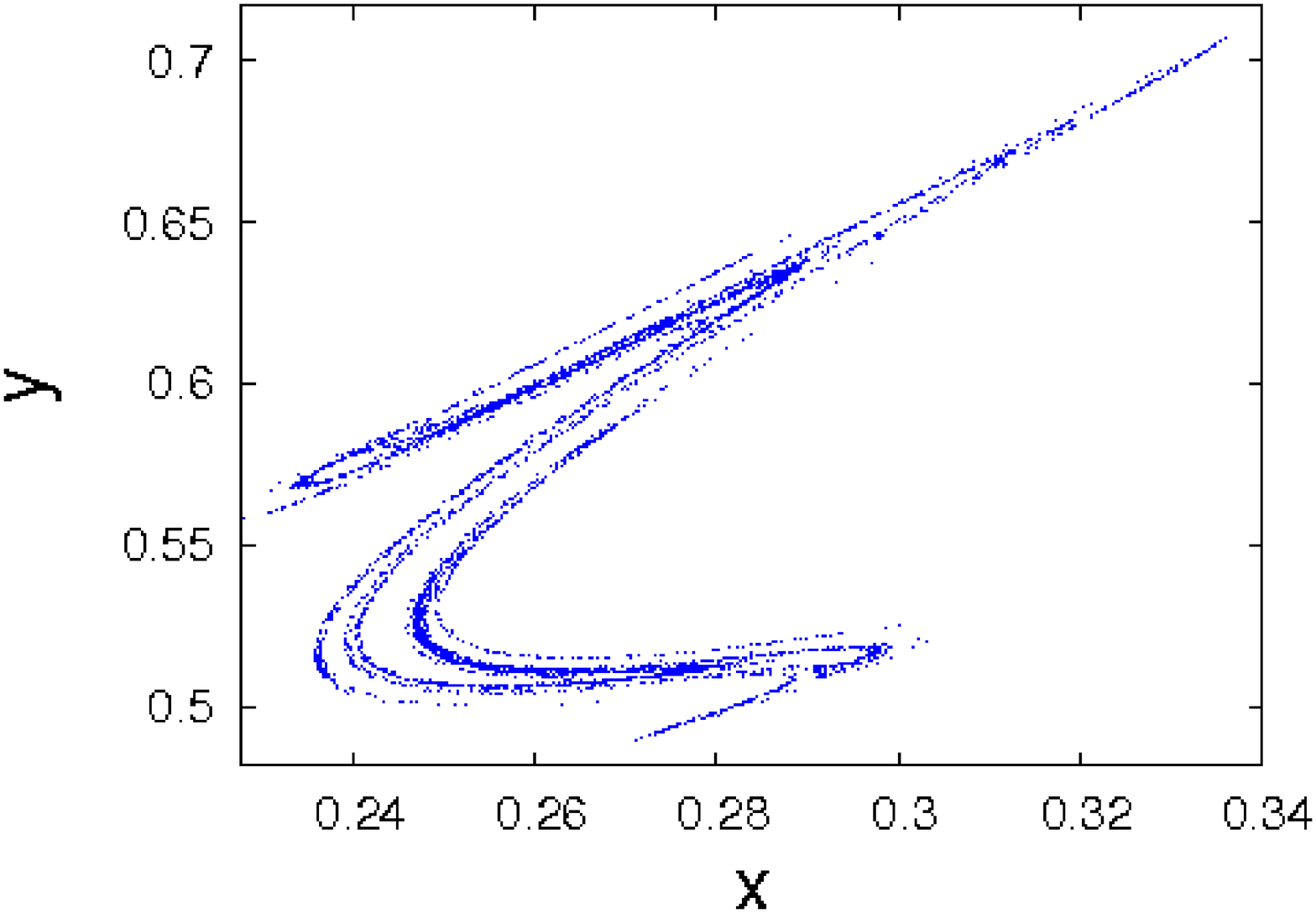} &
\includegraphics[height=1.7in,width=2.1in]{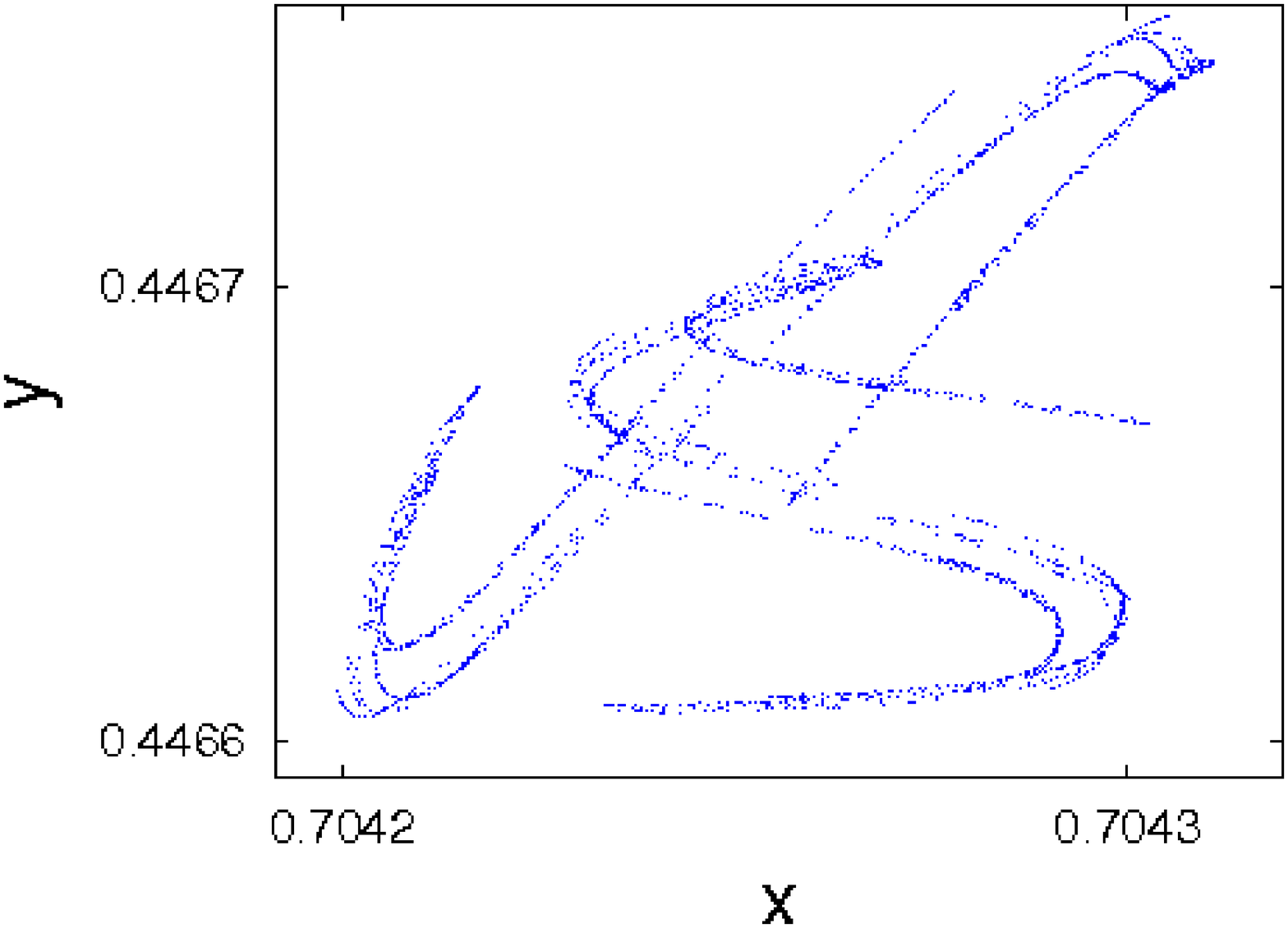} \\
\mbox{(a)} & \mbox{(b)}  \\[0.2cm]
\includegraphics[height=1.7in,width=2.05in]{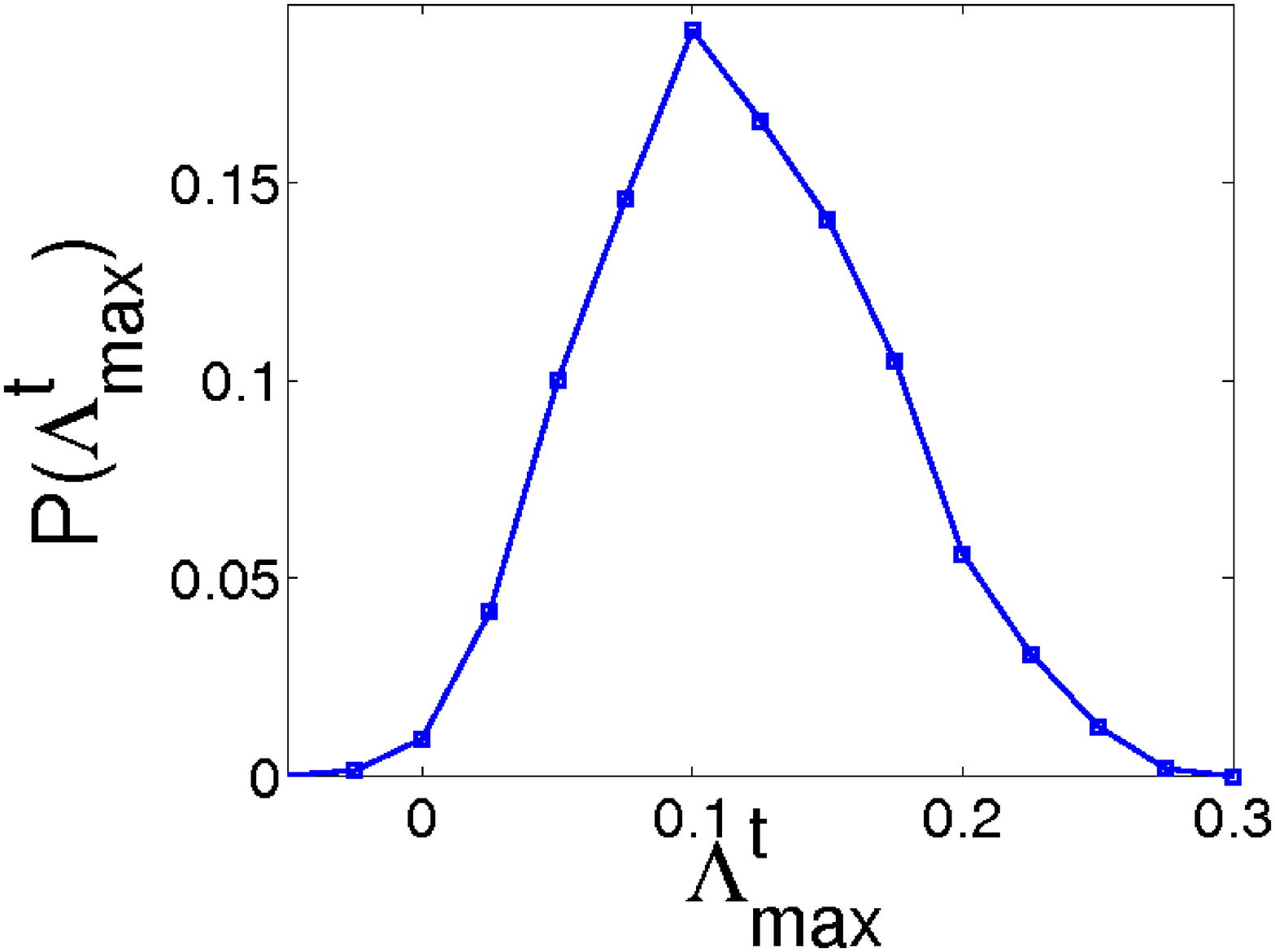} & 
\includegraphics[height=1.7in,width=2.05in]{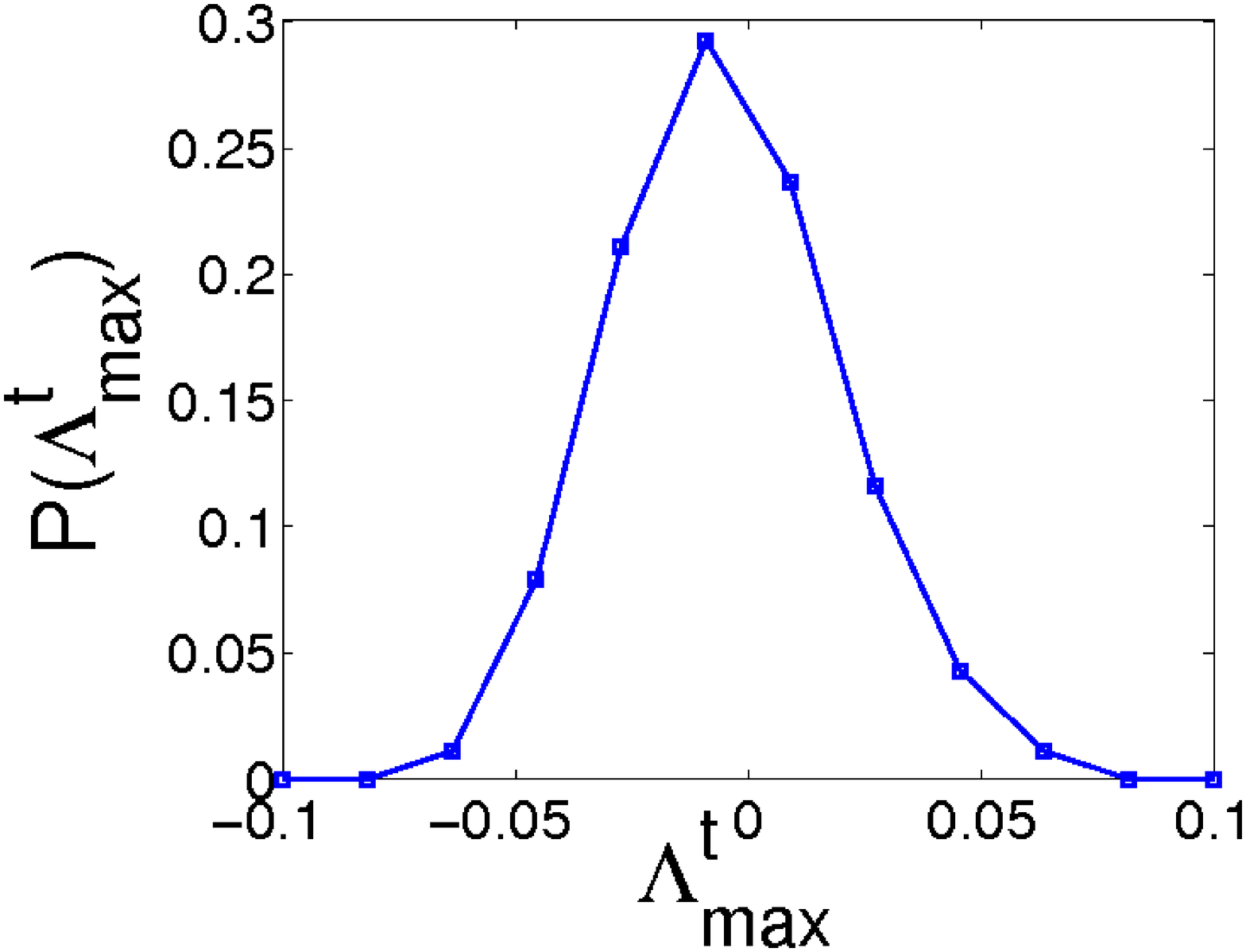} \\
\mbox{(c)} & \mbox{(d)} 
\end{array}$ 
\caption{Examples of strange attractors with (a) $\lambda_{max}^t=0.118$ and (b) $\lambda_{max}^t=-0.005$, both for a branch-node on the 4-star. (c) and (d) Spectrum of $\Lambda_{max}^t(\x_0)$ values along the trajectories shown on (a) and (b), respectively.}
       \label{attractors}
\end{center}
\end{figure}

The observed excessive number of non-periodic orbits, as compared to the orbits with an average $\lambda_{max}^t >0$, suggests that we take a close-up look at the properties along the non-periodic orbits. The two examples of such orbits found for the branch-node on the 4-star for $\mu=0.048$ end-up at the strange attractors, parts of which are  shown in Fig.\,\ref{attractors}a\&b. 
A systematic study of the exponent $\Lambda_{max}^t (\x_0)$, defined in Eq.\ (\ref{mledefinition}), at different points $\x_0$ along the trajectory gives the spectrum of $\Lambda_{max}^t$ values. For the two trajectories in Fig.\,\ref{attractors}a\&b the respective spectra are given in Fig.\,\ref{attractors}c\&d. Note that the  $\lambda_{max}^t$, reported above, roughly corresponds to the peak value of the spectrum. The fractal attractors, as those shown in Fig.\,\ref{attractors}a\&b, have a common feature: the spectra have tails on both positive and negative values of  $\Lambda_{max}^t(\x_0)$.  Specifically, the attractor in Fig.\,\ref{attractors}b has a negative distribution average value $\Lambda_{max}^t$ close to zero. Attractors of this type are referred to as {\it strange non-chaotic attractors} SNA \cite{feudel}, indicating the atypical presence of a negative Lyapunov exponent accompanied with the fractal structure.

\begin{figure}[!hbt]
\begin{center}
$\begin{array}{cc}
\includegraphics[height=1.8in,width=2.5in]{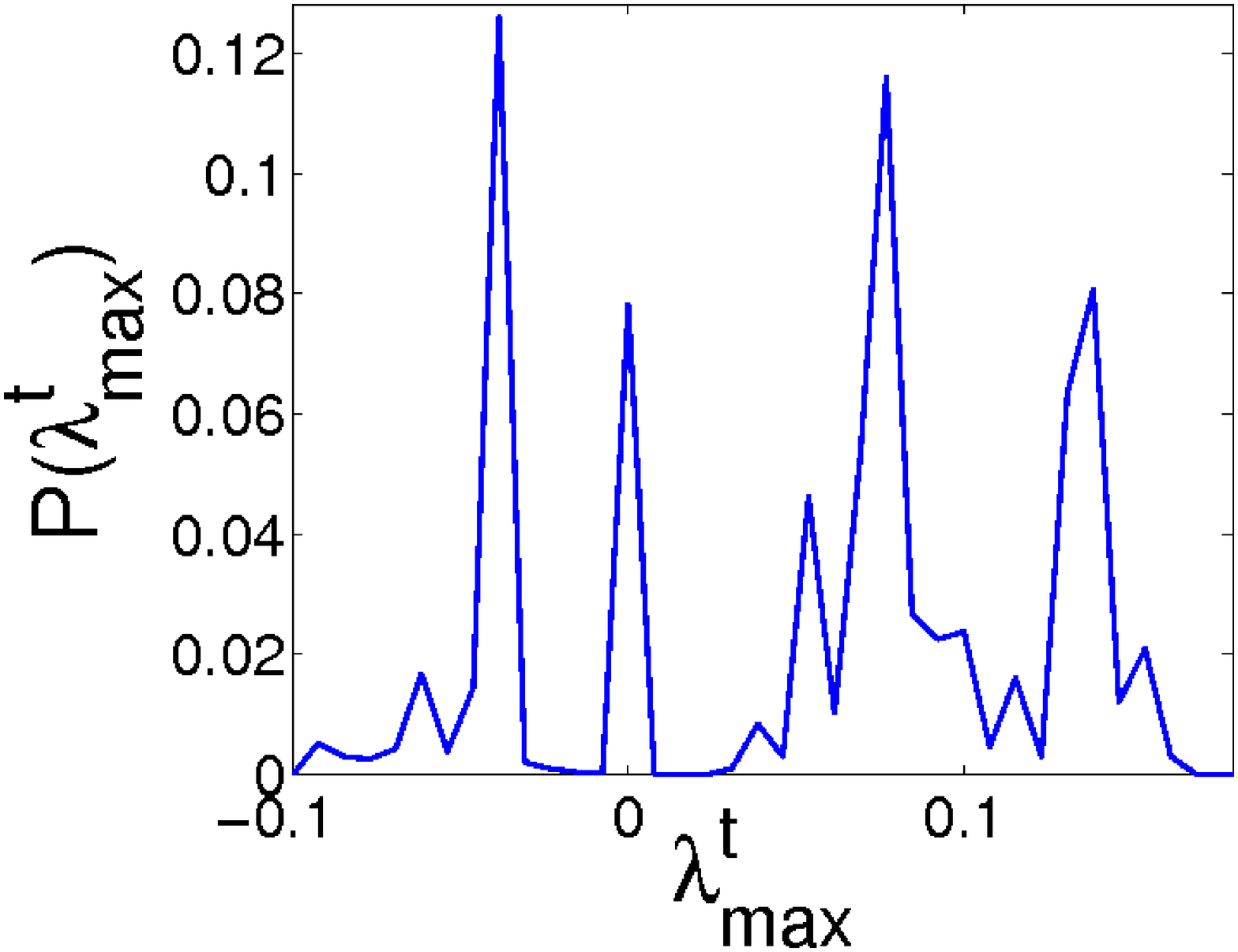} & 
\includegraphics[height=1.8in,width=2.4in]{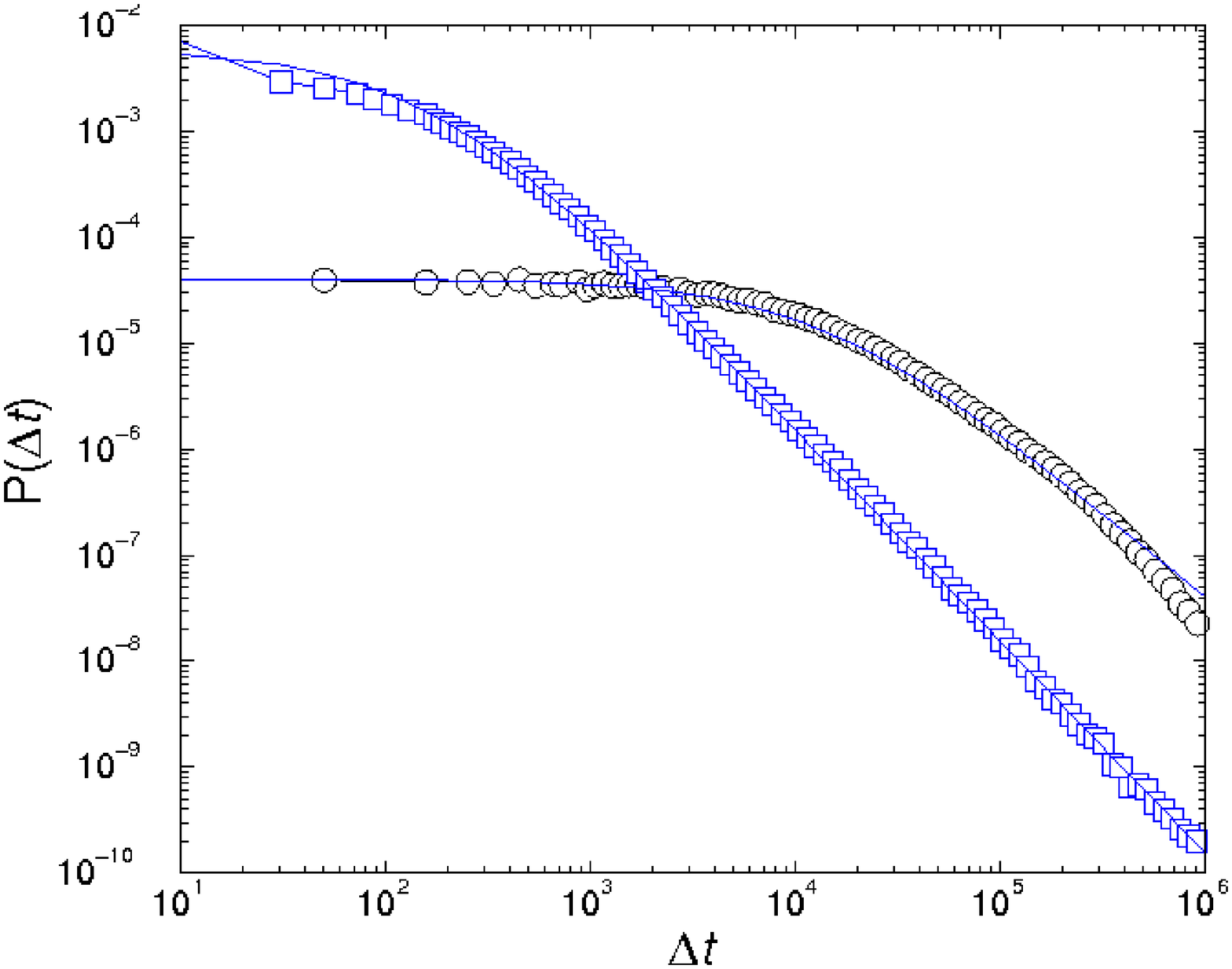}\\
\mbox{(a)} & \mbox{(b)}
\end{array}$ 
\caption{(a) Distribution of $\lambda_{max}^t$ for $10^3$ trajectories of the branch-node on the 4-star for $\mu=0.048$. Averaging was done over spectrum with 100 values of $\Lambda_{max}^t(\x_0)$ computed along each trajectory. (b) Distribution of return times for two trajectories, parts of which are shown in Fig.\,\ref{attractors}a\&b, with  smaller and large slope, respectively.}
\label{048} 
\end{center}
\end{figure}

We find a large number of  strange attractors in our CMS. An incomplete search for fractal attractors gives the distribution of the peak values  $\lambda_{max}^t$ of the respective spectra shown in Fig.\,\ref{048}a, indicating roughly four different groups of such attractors in our CMS.

Further interesting question is how the appearance of such attractors affects the collective motion of the CMS. Here we present the distributions of the return times to a given portion of the phase space, $P(\Delta t)$, defined above. For the two attractors in Fig.\,\ref{attractors}a\&b the results for these  distributions are shown in Fig.\,\ref{048}b. In both cases the distributions of return times exhibit a power-law tail, which can be fitted with the q-exponential function in Eq.\,(\ref{q-exp-form}), suggesting a non-ergodic dynamics near these attractors. Although both attractors in Fig.\,\ref{attractors}a\&b have quantitatively similar fractal dimension, close to 1.5, obviously their other features  affect the return-time statistics. In particular, the larger slope and smaller characteristic scale $X_0=80$ was found in the case of the SNA of Fig.\,\ref{attractors}b, whereas, at the attractor Fig.\,\ref{attractors}a with a positive $\lambda_{max}^t$, we find $X_0=8000$ and an increased probability of long return times. Consequently, the parameter $q$ in the expression Eq.\ (\ref{q-exp-form}), which is a measure of non-ergodicity, appears to be different, i.e., $q=1.6$ for the attractor in Fig.\,\ref{attractors}a and $q=1.5$ for SNA in Fig.\,\ref{attractors}b.

\section{Conclusion}\label{Conclusions}
We have demonstrated that a variety of new dynamical effects occur in two-dimensional standard map with a non-symplectic coupling and time delay between coupled units on tree-like graphs. By focusing on a single node coupled to the tree, we were able to apply the established methods of the discrete-time dynamics, and thus closely follow how a coupled unit adapts to the evolution of the extended dynamical system.  In particular, as a consequence of the coupling we find the dynamical localization of orbits, emergent periodicity, non-periodic orbits  and occurrence of fractal attractors in a wide range of coupling strengths. The self-organized dynamical behavior of mutually coupled maps in our CMS is characterized by power-law dependencies, specifically  in the emergent (tree averaged) periodicity and in the return time intervals near the strange attractors. 

We have shown that some of the strange attractors in our CMS  have non-positive Lyapunov exponents, similarly to the strange non-chaotic attractors \cite{feudel,rama97,rama99} found in the quasi-periodically forced  maps. It should be stressed that in our two-dimensional coupled maps no external driving was imposed.  Contrary to the forced isolated maps, in our CMS different coupled units on the 4-star or on the scale-free tree, provide the dynamical input to each other. Full understanding of the routes to the strange non-chaotic attractors in this coupled-map system requires additional study. We speculate that both the observed localization of orbits and time delay between coupled units, which pervents full synchronization of neighboring nodes, play an important role.

With the parallel analysis of the dynamics at the 4-star motif and on the large tree, in this work  we obtained certain quantitative arguments, which imply how the self-organized dynamic on the large scale networks occurs. Specifically, the 4-star appears as a basic dynamical unit on which  the non-periodic orbits and strange attractors appear and spread on the entire tree, cf. Fig.\,\ref{periodictransition}. The role of cycles on the graph as well as the effects of different/variable delays  between units are left for a separate work.  Further interesting subjects in the context of our CMS, as the appearance of partial synchronization patterns (see also \cite{ja3}) and the problem of coexisting attractors on the graph, are left for the future study.\\[0.1cm]

\n {\bf Acknowledgments}: This work was supported by the Program P1-0044 of the Ministery of Higher Education, Science and Technology of the Republic of Slovenia. We would like to thank M. \v Suvakov, G. Rodgers and R. Ramaswamy for helpful discussions and comments. BT also thanks for a partial support by the COST P10 action ``Physics of Risk''.

\bibliographystyle{ieeetr}
\bibliography{references.bib}

\end{document}